\documentclass[preprint,1p,compress]{elsarticle}  

\usepackage{a4wide,epsfig,amsmath,latexsym}
\usepackage{multirow}
\usepackage{axodraw}

\usepackage{color,graphicx}

\usepackage{graphicx}

\newcommand{\promille}{%
  \relax\ifmmode\promillezeichen
        \else\leavevmode\(\mathsurround=0pt\promillezeichen\)\fi}

\newcommand{\promillezeichen}{%
  \kern-.05em%
  \raise.5ex\hbox{\the\scriptfont0 0}%
  \kern-.15em/\kern-.15em%
  \lower.25ex\hbox{\the\scriptfont0 00}}

\newcommand{\nn}{ \nonumber }  

\newcommand{\al}{\alpha}
\newcommand{\be}{\beta}

\newcommand{\ep}{\varepsilon}

\newcommand{\ice}[1]{\relax}

\newcommand{\pivo}[1]{\relax}

\newcommand{\beq}{\begin{equation}}
\newcommand{\eeq}{\end{equation}}
\newcommand{\bea}{\begin{eqnarray}}
\newcommand{\eea}{\end{eqnarray}}

\newcommand{\ba}{\begin{array}} 
\newcommand{\ea}{\end{array}} 
 
\newcommand{\als}{\alpha_s}

\newcommand{\Ga}{\Gamma}

\newcommand{\g}{\gamma}

\newcommand{\msbar}{\overline{\mbox{MS}}}

\newcommand{\re}[1]{(\ref{#1})}

\newcommand{\sbz}{ }

\newcommand{\unl}[1]{\underline{#1}}

\def\bbuildrel#1_#2^#3%
{\mathrel{\mathop{\kern 0pt#1}\limits_{#2}^{#3}}}

\begin{document}

\begin{frontmatter}
\title{
\centerline{\normalsize\hfill TTP10-18}
\centerline{\normalsize\hfill SFB/CPP-10-24}
\vspace{1.5cm}
Four Loop Massless Propagators: 
an Algebraic  Evaluation of All 
Master Integrals\tnoteref{SGG}
}
\tnotetext[SGG]{In memoriam Sergei Grigorievich Corishny, 1958-1988}

\author[mgu]{P.~A.~Baikov}
\ead{baikov@theory.sinp.msu.ru}

\author[ttp,inr]{K.~G.~Chetyrkin\corref{cor1}}
\cortext[cor1]{corresponding author}
\ead{konstantin.chetyrkin@kit.edu}

\address[mgu]{Skobeltsyn Institute of Nuclear Physics, Moscow State 
University, Moscow~119991, Russia}

\address[ttp]{ Institut f{\"u}r Theoretische Teilchenphysik,
            Karlsruhe Institute of Technology (KIT),
            D-76128 Karlsruhe, Germany}

\address[inr]{Institute
    for Nuclear Research, Russian Academy of Sciences, Moscow 117312, Russia.}

\ice{
}

\begin{abstract}

The old "glue--and--cut" symmetry of massless propagators, first
established  in \mbox{Ref. \cite{Chetyrkin:1981qh}}, leads --- 
{\em after reduction to  master integrals
is performed }--- to a host of
non-trivial  relations
between the latter.  
The relations  constrain the master integrals so tightly that  they all  can
 be analytically expressed in terms of only few, essentially
trivial, watermelon-like integrals. As a
consequence 
we arrive at  explicit analytical results for all master
integrals appearing in the process of reduction of massless
propagators  at three and four loops. 
The transcendental structure of the results suggests  a clean  explanation  
of the well-known  mystery  of the absence of   even zetas ($\zeta_{2n}$)  
in the  Adler function and other similar functions essentially reducible
to  massless propagators. Once a reduction of massless propagators at five loops is  available,  our 
approach should be also applicable for explicitly  performing the corresponding  {\em five-loop master
integrals}. 

\end{abstract}

\end{frontmatter}

\newpage

\tableofcontents

\section{Introduction  \label{intro}}


Within perturbation theory quantum-theoretical amplitudes are
described by Feynman Integrals (FI's). The evaluation of the latter has seen
quite a lot of progress during last three decades. In fact, it has been
elevated  from a collection of loosely related prescriptions
to a solid part of mathematical physics as was recently certified by
the appearance of Smirnov's bestseller books ``Evaluating Feynman integrals"
and (even!) ``Feynman integral calculus'' 
 \cite{Smirnov:2004ym,Smirnov:2006ry}.

A significant number of higher order calculations are performed
according to the following ``standard'' scenario. First, the 
Feynman amplitudes are  reduced to a limited set of 
so-called {\em master integrals} (MI's).
The particular way of implementing the reduction 
is not  unique and not essential  for our discussion\footnote{
The  so-called Laporta approach \cite{Laporta:1996mq,Laporta:2001dd,Gehrmann:1999as} 
seems to be  most often utilized  
but a few other promising methods are  being now  actively developed  
\protect\cite{Baikov:tadpoles:96,Baikov:1996cd,Baikov:explit_solutions:97,Tarasov:1998nx,Smirnov:2005ky,Lee:2008tj}.}.
At the second and final step the resulting master integrals should be computed. 

An important feature of the standard scenario is that the resulting
set of master integrals should be computed only once and forever due
to the well-established\footnote{At least well-established in practice.
See below for an instructive particular example of a class of massless
propagators and also \cite{Baikov:criterion:00,Baikov:2005nv} for an
attempt to formalize the concept of the masters integrals and to prove
the universality property in general. A related discussion could be
found in \cite{Smirnov:2003kc,Smirnov:2006wh,Smirnov:2007iw,Smirnov:2008iw}.}
property of universality: for every
given class of Feynman amplitudes characterized by the number of loops
and the pattern of external momenta and masses 
the corresponding set of master integrals is universal in the following
sense:

(a) Every (even extremely complicated)  amplitude from the class can be
expressed in terms of  one and the same (finite!) set of masters
integrals. 

(b)  The knowledge of MI's up to some  properly fixed order in the
$\ep$-expansion is enough to calculate the finite  part of the amplitude.  
Let us consider an L-loop integral $P$. The reduction to masters
leads to an  identity of the form:
\beq
P = \sum_i C_i(\ep= 2 - D/2)\, M_i
\label{reduction}
{},
\eeq
where sum goes over all relevant master integrals and $C_i(\ep)$ is 
a {\em rational} function of the  space time
dimension $D=4-2\ep$ and kinematical parameters likes masses,
external momenta, etc.  The functions could be singular at the point
$D=4$. The corresponding poles in $\ep$ are customarily referred to as
{\em spurious} ones.  While the coefficients $C_i(\ep)$ depend, obviously, on the
initial integral $P$,  the {\em maximal powers}, $p_i$, of the spurious poles inside  
a given  $C_i(\ep)$ depend {\em  only} on the choice of the basis of master
integrals\footnote{It was proven in \cite{Chetyrkin:2006dh} that  
there  always exists such a set  master integrals that {\em all} coefficient functions 
will be regular at $\ep$ around  zero.}.


Thus, within the standard scenario,  to evaluate  an L-loop amplitude $F$ 
one   proceeds in  three main steps:

\noindent
({\it i}) Choose a set  of master integrals.

\noindent
({\it ii}) Reduce every Feynman integral contributing to the amplitude
$F$ to form \re{reduction}.

\noindent
({\it iii}) Compute the $ \ep \rightarrow 0$  expansion of each master integral $M_i$
up to  (and including) the  term of order  $\ep^{p_i}$.



The steps (i) and (ii) are, in fact, strongly interrelated. In
(almost) all approaches to reduction one first tries to use the
traditional method of Integration By Part (IBP) identities\footnote{
In addition to the IBP identities the  so-called Lorentz-invariance
ones \cite{Gehrmann:1999as} are also often employed in practical
calculations. In fact, the second set of identities has been proved
\cite{Lee:2008tj} to be a consequence of the first one. } in order to
reduce (read simplify) initial integrals as much as possible.  The
remaining basis set of further irreducible (at least in practice)
integrals is considered as the set of MI's. 
As  this final set is usually  rather small it is not of any  practical importance whether  the
corresponding integrals are really independent or not\footnote{In addition,  
sometimes there are  {\em implicit} confirmations of the independence. For instance, if one computes
a gauge invariant combination  of Feynman integrals, then the gauge  independence of a coefficient function
of a MI could be  only guaranteed if the latter is independent from all the others, see, e.g. 
\protect\cite{Schroder:2002re}.}.

Once the set of MI's $M_i$ is fixed, then the corresponding powers $p_i$
can be easily read off  from  the results of reduction of some test set of initial FI's.
Of course if a set of input FI's is too limited, it might happen that a in few cases
an "experimentally" determined power $p_i$ will be smaller  than  its true value. 
Luckily, the basis set of MI's (together with corresponding 
maximal values of spurious poles in their coefficient functions) is usually 
determined in early stage after  calculation  of relatively small subset of all FI's 
to be computed.

The choice of MI's is not unique.  One of the basic criteria is
  simplicity of the calculation of MI's. For example, in view of an
analytical evaluation it is natural to seek for MI's with minimal
number of propagators. On the other hand, for a numerical evaluation
it is often advisable to consider more complicated but less singular
MI's (see, e.g. \cite{Chetyrkin:2006dh}).

\begin{figure}[th]
\thicklines
\begin{center}

\begin{picture}(65,80) 
\put(25,0){$T_1,\ep^2$}
\CArc(36,30)(15,0,360)
\Line(20,30)(15,30)
\Line(20,30)(57,30)
\end{picture}
\begin{picture}(75,60) 
\put(29,0){$T_2,\ep$}
\CArc(33,30)(10,0,360)
\CArc(52.5,30)(10,0,360)
\Line(13,30)(22,30)
\Line(62,30)(72,30)
\end{picture}

\begin{picture}(55,55) 
\put(15,0){$N_0,\ep^0$}
\CArc(26,30)(15,0,360)
\Line(15,41)(37,19)
\Line(37,41)(28,32)
\Line(15,19)(23.4852813742386,27.4852813742386)
\Line(10,30)(5,30)
\Line(42,30)(47,30)
\end{picture}
\begin{picture}(65,80) 
\put(20,0){$L_1,\ep^2$}
\CArc(25,30)(15,0,360)
\CArc(35,30)(15,0,360)
\Line(9,30)(4,30)
\Line(51,30)(56,30)
\end{picture}
\begin{picture}(65,80) 
\put(15,0){$P_1,\ep^3$}
\CArc(26,30)(15,0,360)
%
\Line(26,45.1)(42,30)
\Line(26,45.1)(10,30)
\Line(10,30)(5,30)
\Line(42,30)(47,30)
\end{picture}
\begin{picture}(65,80) 
\put(15,0){$P_2,\ep^3$}
\CArc(10,30)(10,0,360)
\CArc(36,30)(15,0,360)
\Line(0,30)(-5,30)
\Line(20,30)(57,30)
\end{picture}
\begin{picture}(45,80) 
\put(10,0){$P_3,\ep^4$}
\CArc(20,34)(11,0,360)
\CArc(20,26)(11,0,360)
\Line(9,30)(4,30)
\Line(31,30)(36,30)
\end{picture}
\begin{picture}(75,80) 
\put(29,0){$P_4,\ep^2$}
\CArc(18,30)(7.5,0,360)
\CArc(33,30)(7.5,0,360)
\CArc(48,30)(7.5,0,360)
\Line(10,30)(5,30)
\Line(56,30)(61,30)
\end{picture}
\end{center}
\caption{two- and three-loop master p-integrals. $\ep^m$ after a master label  stands for the maximal term in
$\ep$-expansion of the master integral  which one needs to know for  
evaluation of the contribution of the integral to the final  result.
}
\label{3:loop:masters}
\end{figure}
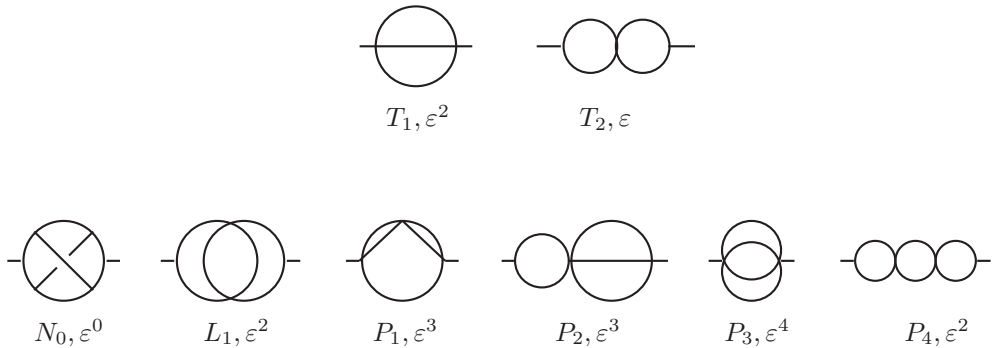

The standard scenario was first developed for  massless propagators
\cite{Chetyrkin:1980pr,Tkachov:1981wb,Chetyrkin:1981qh}. It is no wonder that 
our understanding of reduction and MI's is most advanced for this
case. Indeed, at three-loop level there is an explicit algorithm of
reduction \cite{Chetyrkin:1981qh}  to  
MI's (see, Fig.~\ref{3:loop:masters}).
The existence of such an algorithm proves (a)-universality while
the (rather tedious) analysis of the structure of the algorithm
demonstrates that (b)-universality is also valid
\cite{Chetyrkin:1981qh}.  

\SetWidth{.8}

Let us consider the next loop level in the same class, that is four-loop
massless propagators.  Here the full set of {\em independent} MI's was
theoretically constructed in \cite{Baikov:2005nv}. Then  a special
procedure of reduction, based on $1/D$ expansion of the coefficient
functions of MI's was developed  by one of the present authors
\cite{Baikov:2005nv,Baikov:2007zza} with the help of a
special parametric representation of FI's, elaborated in
\cite{Baikov:tadpoles:96,Baikov:1996cd,Baikov:explit_solutions:97}. 
The $1/D$ method of reduction has been heavily exploited in a
series of publications 
\cite{Baikov:2001aa,Baikov:2004tk,Baikov:2005rw,Baikov:2006ch,Baikov:2008jh,Baikov:2009bg,PhysRevLett.104.132004}
in order to compute a number of
important physical observables in pQCD. { We can not go
here into the  technical details of the four-loop reduction except for
the one: it requires huge computer storage resources and their effective
management. As a consequence its practical implementation would hardly
be feasible without excellent possibilities for dealing with gigantic
data streams offered by the computer algebra language FORM \cite{Vermaseren:2000nd} and, especially,  its versions 
ParFORM  \cite{Fliegner:2000uy,Tentyukov:2004hz,Tentyukov:2006pr} 
and TFORM \cite{Tentyukov:2007mu}}.

Thus, we conclude that the reduction problem for the four-loop
massless propagators is solved in the practical sense.  Analytical
evaluation of the corresponding MI's is the central theme of the
present work.

The plan of the paper is as follows.
Next two sections provide the reader with general information about the problem. Section \ref{3loop} 
explains the essence of our approach in detail on the (now easy) example of
three-loop master integrals.  The really new results are described in
section  \ref{4loop}. It is there  we send an expert in multiloop  calculations directly. 
Section \ref{perspectives} discusses perspectives of our method as for
its extensions to  more loops and other kinematical situations.
In section \ref{evenz}   we  demonstrate   some  peculiar properties of our results
which help to solve an  old puzzle of absence of even zetas from 
some quantities, like the Adler function,  expressible in terms of 
p-integrals.  A discussion of our results is put in section  \ref{discussion}. 
In the last section \ref{conclusions} we summarize the content of  the  paper
and express our gratitude  to people and organizations who (which)
have  been continuously supporting us during the painfully  long period of preparation of the
present publication.

Our results for all four-loop MI's (together with some auxiliary
information)  are  available (in computer-readable form) in
\\
{\tt
http://www-ttp.physik.uni-karlsruhe.de/Progdata/ttp10/ttp10-18.
}

\section{Massless Propagators \label{massless:props}}

Propagators --- that is Feynman integrals depending on only one
external momentum --- appeared in Quantum Field Theory  from its very origin
and since then form
an important class  of FI's.   
Within perturbation theory, every two-point Green function
\beq
G(q) = \int \mathrm{d}x \, e^{iqx}\, \hat{G}(x), 
 \ \ \ \hat{G}(x) \equiv \,\langle 0|T\left[ j_2(x) \, j_1(0) \right] |0\rangle 
{},
\label{G:2point}
\eeq
with $j_1$ and $j_2$ being in general either elementary fields or
(local) composite operators, is expressed in terms of propagators.  If
the momentum transfer $q$ is considered as large with respect to all
relevant masses,  the propagators contributing to $G(q)$ can be
effectively considered as massless.
In what follows we will customarily refer to massless propagator-type FI's
as  {\em  $p$-integrals}. 

p-Integrals appear in many important physical applications.  
Below we briefly mention some most known/important  ones (for more details and examples
see,  e.g. reviews \cite{Chetyrkin:1996ia,Steinhauser:2002rq}).

\begin{itemize}

\item 
The total cross-section of $e^+ e^- $ annihilation into hadrons, the
Higgs decay rate into hadrons, the semihadronic  decay rate of the
$\tau$ lepton and the running of the fine structure coupling are all
computable in the high energy limit in terms of p-integrals. This is
because these quantities are either defined in terms of a  two-point
function \re{G:2point} with properly chosen currents $j_1$ and
$j_2$ 
or can be reduced to this form via  the optical
theorem. 

Note,  that by  high-energy limit we understand not
only the case when all  masses can be neglected but also  the   possibility 
to take into account mass effects by exploiting a small mass  expansion.
As  a suitable example one could mention the calculation
of the power suppressed (of order $m_q^2/s$, $m_q^4/s^2$ and so on) corrections for
the correlators of (axial)vector quark currents in higher orders of pQCD 
\cite{ChetKuhn90,Chetyrkin:1997qi,Chetyrkin:2000zk,Baikov:2004ku,Baikov:2009uw}.

\item Coefficient functions of  short distance Operator Product Expansion
(OPE) of two composite operators can  be always expressed in terms of
p-integrals with the help of so-called method of projectors
\cite{Gorishnii:1983su,Gorishnii:1986gn}.  A good example of an early
multiloop OPE calculation is the one of the $\als^3$ corrections to
the Bjorken sum rule for polarized electroproduction and to the
Gross-Llewellyn Smith sum rule \cite{Larin:1991tj}.


\item 
p-Integrals are extremely useful in Renormalization Group (RG) calculations
within the framework of Dimensional Regularization \cite{Ashmore:1972uj,Cicuta:1972jf,'tHooft:1972fi} and 
Minimal Subtractions (MS) schemes \cite{'tHooft:1973mm}. 
\end{itemize}

The naturalness and convenience of the MS-scheme for RG calculations
comes from the following statement \cite{Collins:1974da}:

\vglue 0.2cm
{\bf Theorem   1.} {\it Any UV counterterm for any FI 
integral and, consequently, any RG function in an  arbitrary minimally
renormalized model is a polynomial in momenta and masses}.
\vglue 0.1cm

\vglue 0.1cm

\noindent
This  observation was effectively employed by
A. Vladimirov \cite{Vladimirov:1979zm}
to simplify considerably  the calculation of the  RG
functions.  The  method  was further  developed and named
Infrared Rearrangement (IRR) in \cite{Chetyrkin:1980pr}. It essentially amounts to
an appropriate transformation of the IR structure of  FI's by setting
zero some external momenta and masses (in some cases after 
the differentiation is performed with respect to the latter).
As a result  the calculation of  UV counterterms is much simplified
by reducing the problem to evaluating  $p$-integrals.  
The method of IRR was ultimately refined
and freed from unessential complications   by inventing a so-called
$R^*$-operation \cite{Chetyrkin:1982nn,Chetyrkin:1984xa}. 
The  main use of the $R^*$ -operation
is in the proof of the following statement \cite{Chetyrkin:1984xa}:
\vglue 0.2cm

\noindent
{\bf Theorem 2.} {\it Any (L+1)-loop UV counterterm for any
Feynman integral may be expressed in terms of pole and finite parts
of some appropriately  constructed L-loop  $p$-integrals.}
\vglue 0.1cm

Theorem  2  is a key tool  for  multiloop RG
calculations as it reduces  the general task of evaluation of
(L+1)-loop  UV counterterms to a well-defined and clearly posed
purely mathematical problem: the calculation of L-loop $p$-integrals.
In the following  we shall refer  to the latter as the  L-loop Problem.

The one-loop Problem is  trivial (see eq. \re{1-loop:generic} in the next section).
The two-loop Problem was  solved after inventing and developing the 
Gegenbauer polynomial technique in $x$-space (GPTX)  \cite{Chetyrkin:1980pr}.  {\em In
principle} GTPX is  applicable to compute  analytically some quite non-trivial
three and even higher loop  p-integrals\footnote{
The GPTX is also ideally suited for high-precision numerical
calculations  of {\em finite} p-integrals ({\em  with simple or, better,  without numerators})
with really many  loops. 
See \cite{Broadhurst:1995km,Fiamberti:2008sh,Fiamberti:2009jw} for a number of spectacular examples in four, five, six and even seven loops.
} (for  a review see \cite{Kotikov:2001sd}).
However, in
practice calculations quickly get clumsy, especially for diagrams with
numerators.  The main breakthrough at the three-loop level happened with
elaborating the method of integration by
parts  \cite{Tkachov:1981wb,Chetyrkin:1981qh} of dimensionally
regularized integrals. All (about a dozen) topologically different
families of three-loop $p$-integrals were neatly analyzed in \cite{Chetyrkin:1981qh} and a
explicit calculational algorithm was suggested for every case. As
a result the algorithm of integration by parts for three-loop
$p$-integrals was established. Later the algorithm was implemented
(and named MINCER) within the computer algebra languages SCHOONSCHIP
\cite{Strubbe:1974vj} and FORM \cite{Vermaseren:2000nd} (see
Refs~\cite{Gorishnii:1989gt} and \cite{Larin:1991fz} respectively).
The most recent  FORM  version of MINCER is freely available 
from  {\verb+ http://www.nikhef.nl/~form+}.

During last two  decades  MINCER has  been  used intensively  to
perform a number of important calculations of higher order radiative
corrections in various field theories. As a couple of outstanding examples, characterizing  the issue,
    we  mention the   analytical evaluation of the ${\cal O}(\alpha_s^3)$ 
correction to the ratio $R $ in massless QCD
\cite{GorKatLar91:R(s):4l,SurSam91}  and  recent  analytical calculations  of three-loop deep-inelastic
structure functions\footnote{These calculations, in fact, have
required development and application of a number of additional
technical tools (including highly advanced version of integration by
parts algorithm) than just the  use of MINCER and the method of
projectors; please consult the original works for further details.}
\cite{Moch:2004pa,Vogt:2004mw,Vermaseren:2005qc}.

Note,  that every L-loop Problem is naturally decomposed  in two: 
(A) reduction of
a generic L-loop p-integrals  to masters   and
(B) evaluation of the latter.   
As   A-problem   has already been discussed, 
we proceed now to   B-problem.  For L equal to 1 or 2 problem B degenerates to a
trivial one due to the fact that all masters, being primitive ones,
are easily evaluated  in terms of $\Gamma$-functions.
At three-loop level there exist only two non-trivial\footnote{More
precisely, we mean non-primitive p-integrals, see definitions below in
section~\ref{G-functions}.} master integrals whose evaluation was rather simple with
the help of GPTX and, in fact, was performed well before the algorithm
of reduction of three-loop p-integrals  was discovered. 
Thus, in  three-loops  A and B  problems  could be considered as 
two  separate ones.

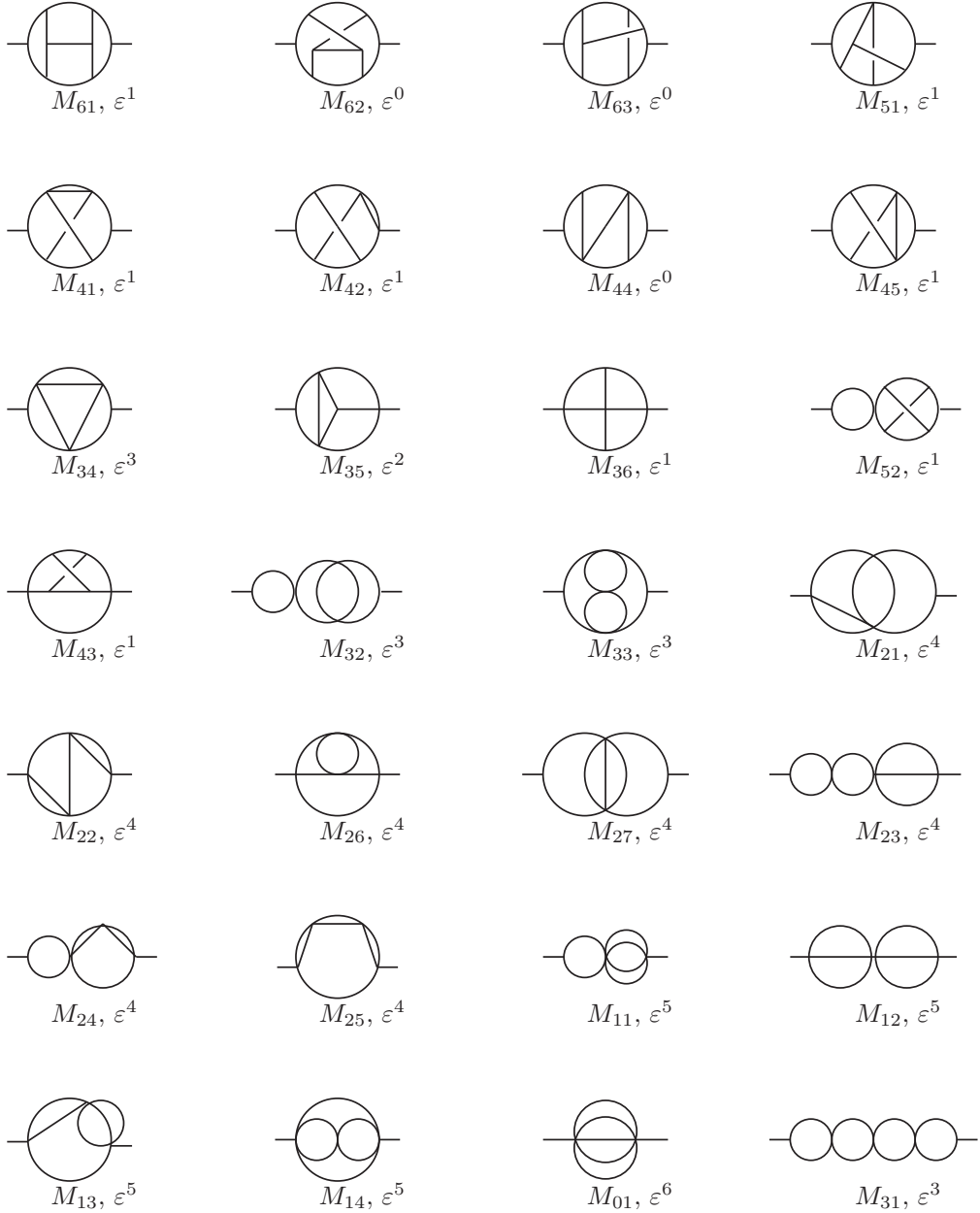
\begin{figure}[!hbt]
\begin{picture}(100,90)(0,0)
\CArc(50,50)(20,0,360)
\Line(70,50)(80,50)
\Line(30,50)(20,50)
\Line(61,67)(61,33)
\Line(39,67)(39,34)
\Line(39,50)(61,50)
\put(33,15){$M_{61},\,\ep^1$}
\end{picture}
\begin{picture}(100,90)(0,0)
\CArc(50,50)(20,0,360)
\Line(70,50)(80,50)
\Line(30,50)(20,50)
\Line(62,34)(62,47)
\Line(38,34)(38,47)
\Line(38,47)(62,47)
\Line(62,47)(36,64.3333333333333)
\Line(38,47)(46.3205029433784,52.5470019622523)
\Line(64,64)(53,56.6666666666667)
\put(33,15){$M_{62},\,\ep^0$}
\end{picture}
\begin{picture}(100,90)(0,0)
\CArc(50,50)(20,0,360)
\Line(70,50)(80,50)
\Line(30,50)(20,50)
\Line(39,67)(39,33)
\Line(39,50)(68,57.25)
\Line(61,33)(61,53)
\Line(61,67)(61,58)
\put(33,15){$M_{63},\,\ep^0$}
\end{picture}
\begin{picture}(100,90)(0,0)
\CArc(50,50)(20,0,360)
\Line(70,50)(80,50)
\Line(30,50)(20,50)
\Line(50,70)(34,38)
\Line(40,50)(65,37.5)
\Line(50,70)(50,48)
\Line(50,30)(50,42)
\put(33,15){$M_{51},\,\ep^1$}
\end{picture}

\vskip -2.em
\begin{picture}(100,90)(0,0)
\CArc(50,50)(20,0,360)
\Line(61,67)(52,53.5)
\Line(39,34)(48,47.5)
\Line(50,50)(39,66.5)
\Line(50,50)(61,33.5)
\Line(61,67)(39,67)
\Line(70,48)(80,48)
\Line(30,48)(20,48)
\put(33,15){$M_{41},\,\ep^1$}
\end{picture}
\begin{picture}(100,90)(0,0)
\CArc(50,50)(20,0,360)
\Line(61,66)(52,52.5)
\Line(39,34)(48,47.5)
\Line(50,50)(39,66.5)
\Line(50,50)(61,33.5)
\Line(61,66)(70,48)
\Line(70,48)(80,48)
\Line(30,48)(20,48)
\put(33,15){$M_{42},\,\ep^1$}
\end{picture}
\begin{picture}(100,90)(0,0)
\CArc(50,50)(20,0,360)
\Line(50,50)(61,66.5)
\Line(50,50)(39,33.5)
\Line(70,48)(80,48)
\Line(30,48)(20,48)
\Line(61,67)(61,33)
\Line(39,67)(39,34)
\put(33,15){$M_{44},\,\ep^0$}
\end{picture}
\begin{picture}(100,90)(0,0)
\CArc(50,50)(20,0,360)
\Line(61,66)(52,52.5)
\Line(39,34)(48,47.5)
\Line(50,50)(39,66.5)
\Line(50,50)(61,33.5)
\Line(61,67)(61,34)
\Line(70,48)(80,48)
\Line(30,48)(20,48)
\put(33,15){$M_{45},\,\ep^1$}
\end{picture}

\vskip -2.0em
\begin{picture}(100,90)(0,0)
\Line(30,50)(20,50)
\Line(70,50)(80,50)
\CArc(50,50)(20,0,360)
\Line(34,62)(66,62)
\Line(50,30)(34,62)
\Line(50,30)(66,62)
\put(33,15){$M_{34},\,\ep^3$}
\end{picture}
\begin{picture}(100,90)(0,0)
\Line(30,50)(20,50)
\Line(50,50)(80,50)
\CArc(50,50)(20,0,360)
\Line(50,50)(41,68)
\Line(50,50)(41,32)
\Line(41,32)(41,68)
\put(33,15){$M_{35},\,\ep^2$}
\end{picture}
\begin{picture}(100,90)(0,0)
\Line(20,50)(80,50)
\CArc(50,50)(20,0,360)
\Line(50,30)(50,70)
\put(33,15){$M_{36},\,\ep^1$}
\end{picture}
\begin{picture}(100,90)(0,0)
\CArc(40,50)(10,0,360)
\CArc(66,50)(15,0,360)
\Line(55,61)(77,39)
\Line(77,61)(67.54,51.54)
\Line(64.46,48.46)(55,39)
\Line(30,50)(20,50)
\Line(82,50)(92,50)
\put(33,15){$M_{52},\,\ep^1$}
\end{picture}

\vskip -2.0em
\begin{picture}(100,90)(0,0)
\CArc(50,50)(20,0,360)
\Line(20,50)(80,50)
\Line(42,68)(60,50)
\Line(58,68)(51.6496,61.6496)
\Line(47.56,57.56)(40,50)
\put(33,15){$M_{43},\,\ep^1$}
\end{picture}
\begin{picture}(100,90)(0,0)
\CArc(19,50)(10,0,360)
\CArc(45,50)(15,0,360)
\CArc(55,50)(15,0,360)
\Line(9,50)(-1,50)
\Line(71,50)(81,50)
\put(33,15){$M_{32},\,\ep^3$}
\end{picture}
\begin{picture}(100,90)(0,0)
\CArc(50,50)(20,0,360)
\CArc(50,60)(10,0,360)
\CArc(50,40)(10,0,360)
\Line(30,50)(20,50)
\Line(70,50)(80,50)
\put(33,15){$M_{33},\,\ep^3$}
\end{picture}
\begin{picture}(100,90)(0,0)
\CArc(40,50)(20,0,360)
\CArc(60,50)(20,0,360)
\Line(50,33)(20,48)
\Line(20,48)(10,48)
\Line(80,48)(90,48)
\put(33,15){$M_{21},\,\ep^4$}
\end{picture}

\vskip -2.0em
\begin{picture}(100,90)(0,0)
\CArc(50,50)(20,0,360)
\Line(70,50)(80,50)
\Line(30,50)(20,50)
\Line(50,30)(50,70)
\Line(30,50)(50,30)
\Line(70,50)(50,70)
\put(33,15){$M_{22},\,\ep^4$}
\end{picture}
\begin{picture}(100,90)(0,0)
\CArc(50,50)(20,0,360)
\CArc(50,60)(10,0,360)
\Line(20,50)(80,50)
\put(33,15){$M_{26},\,\ep^4$}
\end{picture}
\begin{picture}(100,90)(0,0)
\CArc(40,50)(20,0,360)
\CArc(60,50)(20,0,360)
\Line(50,33)(50,67)
\Line(20,50)(10,50)
\Line(80,50)(90,50)
\put(33,15){$M_{27},\,\ep^4$}
\end{picture}
\begin{picture}(100,90)(0,0)
\CArc(20,50)(10,0,360)
\CArc(40,50)(10,0,360)
\CArc(66,50)(15,0,360)
\Line(10,50)(0,50)
\Line(50,50)(92,50)
\put(33,15){$M_{23},\,\ep^4$}
\end{picture}

\vskip -2.0em
\begin{picture}(100,90)(0,0)
\CArc(40,50)(10,0,360)
\CArc(66,50)(15,0,360)
\Line(66,66)(82,50)
\Line(66,66)(50,50)
\Line(30,50)(20,50)
\Line(82,50)(92,50)
\put(33,15){$M_{24},\,\ep^4$}
\end{picture}
\begin{picture}(100,90)(0,0)
\CArc(50,50)(20,0,360)
\Line(31,45)(21,45)
\Line(69,45)(79,45)
\Line(38,66)(62,66)
\Line(38,66)(31,45)
\Line(62,66)(69,45)
\put(33,15){$M_{25},\,\ep^4$}
\end{picture}
\begin{picture}(100,90)(0,0)
\CArc(40,50)(10,0,360)
\CArc(60,53)(10,0,360)
\CArc(60,47)(10,0,360)
\Line(30,50)(20,50)
\Line(70,50)(80,50)
\put(33,15){$M_{11},\,\ep^5$}
\end{picture}
\begin{picture}(100,90)(0,0)
\CArc(34,50)(15,0,360)
\CArc(66,50)(15,0,360)
\Line(10,50)(90,50)
\put(33,15){$M_{12},\,\ep^5$}
\end{picture}

\vskip -2.0em
\begin{picture}(100,90)(0,0)
\CArc(50,50)(20,0,360)
\CArc(65,58)(11,0,360)
\Line(58,68)(30,49.3333333333333)
\Line(30,49)(20,49)
\Line(70,47)(80,47)
\put(33,15){$M_{13},\,\ep^5$}
\end{picture}
\begin{picture}(100,90)(0,0)
\CArc(50,50)(20,0,360)
\CArc(40,50)(10,0,360)
\CArc(60,50)(10,0,360)
\Line(30,50)(20,50)
\Line(70,50)(80,50)
\put(33,15){$M_{14},\,\ep^5$}
\end{picture}
\begin{picture}(100,90)(0,0)
\CArc(50,54)(15,0,360)
\CArc(50,46)(15,0,360)
\Line(20,50)(80,50)
\put(33,15){$M_{01},\,\ep^6$}
\end{picture}
\begin{picture}(100,90)(0,0)
\CArc(20,50)(10,0,360)
\CArc(40,50)(10,0,360)
\CArc(60,50)(10,0,360)
\CArc(80,50)(10,0,360)
\Line(10,50)(0,50)
\Line(90,50)(100,50)
\put(33,15){$M_{31},\,\ep^3 $}
\end{picture}
\begin{center}
\end{center}
\caption{all master p-integrals for the four-loop Problem. In $M_{ij}$  the digit $i$  stands for   the number 
of (internal) lines in the integral minus five and $j$ numerates different integrals with the same value of $i$. 
The integrals are ordered  (if  read from left to right and then from top to bottom) according 
to their complexity.  $\ep^{\scriptsize m}$ after    $M_{ij}$ stands for the maximal term in
$\ep$-expansion of $M_{ij}$ which one needs to know for  
evaluation of the contribution of the integral to the final  result for a four-loop p-integral
after reduction is done. In other words, $m$ stands  for the maximal power of a spurious pole $1/\ep^m$ which
could appear  in front of $M_{ij}$  in the process of reduction to masters.
}
\label{all:masters}
\end{figure}

\begin{figure}[!hbt]
\begin{center}
\includegraphics{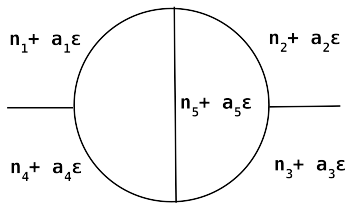}
\end{center}
\caption{the generalized two-loop p-integral; indexes  besides lines show the powers
of corresponding massless propagators. $n_i$ and $a_i$ are assumed to be  integers.}
\label{genF}
\end{figure}

The situation is different in  four loops. In this case  there exist 
\cite{Baikov:2005nv} twenty eight 
master integrals pictured on Fig.~\ref{all:masters} and only 15 of
them (all after $M_{43}$) are simple.  We call a four-loop p-integral
simple if it is either primitive or reducible to the so-called
generalized two-loop F-diagram with insertions, $F(n_1+ a_1 \ep,\dots)$
pictured on Fig.~\ref{genF}.  The corresponding F-integral has been
intensively studied since work 
\cite{Rosner:1967}
and is now in
some sense analytically known \cite{Bierenbaum:2003ud}.  
The remaining 13 masters (from $M_{61}$ till
$M_{43}$,  reading the table from left to right and the from top to
bottom) happen to be quite difficult to deal with even numerically,
not speaking about analytical evaluation.

The aim of the present work is to demonstrate that there exists a
remarkable bootstrap-like connection between parts A and B for the
L-loop Problem irrespectively the specific value of L.
The connection is
powerful enough to result in an explicit solution of problem B for L
equal to three and four (which we will demonstrate explicitly) and, in all
probability, for L equal to five (we will provide the reader with
a strong argument for it).


\thicklines
\SetScale{0.80}
\SetWidth{.8}

The only prerequisite for our considerations is the solution of
problem A  for the corresponding number of loops.

\noindent

We will describe how the  good old "glue--and--cut" symmetry of
massless propagators \cite{Chetyrkin:1980pr} leads --- {\em after
the reduction to  master integrals is performed }--- to
non-trivial relations between the latter.   The relations constraint
the masters integrals so tightly that they can all be analytically
expressed in terms of only few, essentially trivial, watermelon-like
integrals (see diagrams $M_{31}, M_{01},  M_{12}, M_{11}$ and $M_{23}$
on Fig~\ref{all:masters}).  This provide us with explicit analytical
results for all master integrals appearing in the process of reduction
of massless propagators at three and four loops.  By an analytical result we mean, of course, not
an analytical expression for a master integral taken at a generic value
of the  space-time dimension $D$ (which is usually not possible
except for the simplest cases), but rather the one for {\em proper}
number of terms in its Laurent expansion in $D$ around the physical
value $D=4$ as it was discussed in detail above in section \ref{intro}.


Note, that for our aims it is completely irrelevant how exactly the
part A (reduction to masters) is performed/implemented. In fact, we 
only  need the reduction for relatively simple cases of p-integrals:
namely, no squared propagators and relatively low powers of scalar
products in numerators. In particular, no knowledge of (admittedly
rather complicated) reduction techniques based on the asymptotic $1/D$
expansion is necessary. For understanding of all considerations of the
paper it is enough to assume that the reduction (problem A) is
done with some implementation of the Laporta algorithm.

\section{Recursively one-loop integrals \label{G-functions}}

Without loss of generality we will consider the scalar p-integrals
defined in Euclidean space-time.  
Let $F(q,\ep)$ be a dimensionally regulated  scalar $L$-loop  p-integral depending on  
external momentum $q$ and the space-time dimension $D=4-2\,\ep$.  
Its  dependence  on $q$  can be  written as   
\beq
F(q,\ep) = 
f(\ep) \, (q^2)^{\omega/2-L\ep}
\label{generic:p_int}
\eeq
where $\omega$ is the canonical  mass dimension of $F(q,0)$ and
$f(\ep)$ is a meromorphic function of $\ep$.

The complexity of computing of the function of $f(\ep)$ depends 
on the  loop number $L$. At one loop level the result for  
the generic  integral      
\beq
\int \frac{\unl{d} \,\ell } { (\ell^{2\alpha} )(q-\ell)^{2\beta}} =
(q^2)^{2-\ep-\al-\be} G(\alpha,\beta)
\label{1-loop:generic}
\eeq
is known since long (see, e.g.  \cite{Chetyrkin:1980pr}) and reads\footnote{
We provide every loop integration  $d^D l$ with an extra
normalization factor $1/\pi^{D/2}$ and write 
\mbox{$\underline{d}\ell = \frac{d^D\, \ell}{\pi^{D/2}}$}.
}
\beq
 G(\alpha,\beta) = \frac{\Ga(\al +\be-2 +\ep)}{\Ga(\al)\Ga(\be)}
\frac{\Ga(2-\al-\ep)\,\Ga(2-\be-\ep)}{\Ga(4-\al - \be -2\ep)}
\label{G(al,be)}
{}.
\eeq
Here  ``generic'' means that the powers $\al$ and $\beta$
could be not only integers but  functions of $\ep$.
The most useful in applications  case is
\beq
 \al = m +a \ep,  
\ \ \ \ 
 \be = n +b \ep 
\label{al,be}
\eeq
with $n,m$ being arbitrary integers,  $a,b$ 
nonnegative ones.  Note that  {\em negative  values} of 
$a$ and/or $b$ might lead  to  $\ep$ independent singular factor(s) like
$\Ga(0)$ within the corresponding  $G$-function.  On formal grounds 
$G(\al,\be)$ is not defined in this situation\footnote{See, however,
\cite{Gorishnii:1984te} for a significantly deeper discussion of such cases.
  }. 
The reduction formula for G-functions
\beq
G(\al,\be) = \frac{(\al+\be -3+\ep)(4-\al-\be-2\ep)}{(\be-1)(2-\be -\ep)}
G(\al,\be-1)
\label{G:reduction}
\eeq
as well as 
the expansion  
\bea
G(1+a\ep,1+b\ep) &=&  \frac{G_0(\ep)}{\ep( 1+ a+b)}\Big(1+ (a+b)\ep +
(a+b) (a+b+2)\ep^2 + \dots \Big)
{},
\\
G_0(\ep) \equiv  \ep\,G(1, 1) &=&  1 + \ep\,(2  -\g_E) + \dots
\label{G:expansion}
\eea
allows for a convenient evaluation of  
$G(n +a \ep,m +b\ep)$ without any reference to the  awkward formula 
\re{G(al,be)}.
In fact, the well-known freedom  in  the definition of the dimensional
regularization\footnote{The freedom  amounts to the multiplication of every L-loop integral by 
a factor $n(\ep)^L$, with $n(\ep) = 1+ {\cal O}(\ep)$ being a regular (at least in a vicinity of the point $\ep=0$)
function of $\ep$ \cite{Kang:1975rc}.   
Thus,  the formulas \re{G:scheme1}  and \re{G:scheme2} below should be understood 
in the  sense that $n(\ep)$ is chosen as follows
$n(\ep) = 1/(\ep\,G(1,1)) \equiv  {\Gamma(2 -2\ep)}/{(\Gamma(1 +\ep)\Gamma(1-\ep)^2)}$.
\ice{
\[
n(\ep) = \frac{1}{\ep\,G(1,1)} \equiv  \frac{\Gamma(1 +\ep)\Gamma(1-\ep)^2}{\Gamma(2 -2\ep)}
{}.
\]
}
}
allows to tune the function $G_0(\ep)$ at  will 
(provided $G_0(0)=1$). The most natural choice 
\beq
G(1,1)  \equiv \frac{1}{\ep} 
\label{G:scheme1}
\eeq
or, equivalently, 
\beq
G_0(\ep)   \equiv 1
\label{G:scheme2}
\eeq 
fixes the so-called G-scheme \cite{Chetyrkin:1980pr} and will be adopted here.
Note that the  G-scheme is not only extremely convenient from purely 
calculational  point of view; it is also ``natural'' in the realm of 
massless propagators. There is    evidence that 
results expressed in the G-scheme usually tend to display a better pattern 
of ``apparent'' convergence in comparison to the $\msbar$ scheme.


In view of eqs.~\re{generic:p_int} and \re{1-loop:generic} any recursively one-loop p-integral
can be easily performed analytically \cite{Chetyrkin:1980pr}. We will denote  such integrals 
{\em primitive} ones. For example, the two-loop MI's  $T_1$ and $T_2$ (see Fig. 1)
are both  primitive ones, their $\ep$-expansions (with accuracy   necessary for the-two-loop calculation) can be easily computed
via G-functions:
\bea
T_1 &=& 
-\frac{ 1 }{4 \,\ep}-\frac{5}{8 }- \frac{27\ep}{16}+ \,\ep^2 \left( -\frac{153}{32}+\frac{3 \,\zeta_{3} }{2}
\right)+ {\cal O}(\ep^3)
\label{T1}
{},
\\
T_2 &=& \frac{1}{\ep^2} + {\cal O}(\ep^2)
\label{T2}
{},
\eea
with $\zeta_n \equiv \sum_{i \ge 1} \frac{1}{i^n}$. Here and almost everywhere below we set $q^2=1$. 

For future reference we provide  below
expressions in terms of G-functions for the four watermelon-like
primitive three-loop master integrals which serve as building blocks
for all other (three-loop)  masters (see section \ref{3loop}). To make the formulas shorter we
always use the G-scheme defining relation \re{G:scheme1} and write
everywhere $1/\ep$ instead of the   $G(1,1)$:
\begin{align}
P_1 &=  \frac{1}{\ep^2} \, G(2\ep,1),   & P_2 &=  \frac{1}{\ep^2}\, G(\ep,1), 
\nn
\\
P_3 &=  \frac{1}{\ep^2} \, G(\ep,\ep),  & P_4 &=  \frac{1}{\ep^3}
{}.
\label{P1_P4}
\end{align}

\ice{

\section{Three-loop masters integrals: a reminder}


Starting from works \cite{} it was firmly established that there exits
five three-loop master p-integrals pictured on Fig. \ref{fig:}. 
\begin{figure}
\thicklines
{\begin{picture}(55,55) 
\put(15,53){$N_0,\ep^0$}
\put(26,30){\circle{30}}
\put(15,41){\line(1,-1){22}}
\put(37,41){\line(-1,-1){10}}
\put(15,19){\line(1,1){10}}
\put(10,30){\line(-1,0){5}}
\put(42,30){\line(1,0){5}}
\end{picture}
}
\begin{picture}(65,80) 
\put(20,53){$L_1,\ep^2$}
\put(25,30){\circle{30}}
\put(35,30){\circle{30}}
\put(9,30){\line(-1,0){5}}
\put(51,30){\line(1,0){5}}
\end{picture}
\begin{picture}(65,80) 
\put(15,53){$P_1,\ep^3$}
\put(26,30){\circle{30}}
\put(26,46){\line(1,-1){16}}
\put(26,46){\line(-1,-1){16}}
\put(10,30){\line(-1,0){5}}
\put(42,30){\line(1,0){5}}
\end{picture}
\begin{picture}(65,80) 
\put(15,53){$P_2,\ep^3$}
\put(10,30){\circle{20}}
\put(36,30){\circle{30}}
\put(0,30){\line(-1,0){5}}
\put(20,30){\line(1,0){37}}
\end{picture}
\begin{picture}(45,80) 
\put(10,53){$P_3,\ep^4$}
\put(20,34){\circle{22}}
\put(20,26){\circle{22}}
\put(9,30){\line(-1,0){5}}
\put(31,30){\line(1,0){5}}
\end{picture}
\begin{picture}(75,80) 
\put(29,53){$P_4,\ep^2$}
\put(18,30){\circle{15}}
\put(33,30){\circle{15}}
\put(48,30){\circle{15}}
\put(10,30){\line(-1,0){5}}
\put(56,30){\line(1,0){5}}
\end{picture}
\end{figure}
}

\section{Three-loop integrals \label{3loop}}

In this section we discuss the main idea of  our method on a first non-trivial example of three-loop
massless propagator-like integrals.

\subsection{Three-loop finite p-integrals and glueing  \label{3loop:gluieng}}

It is easier to explain the glue-and-cut symmetry on a
real-life example. Almost exactly thirty years ago one of the present
authors was strongly puzzled by the following facts (resulting from
first calculations made with the help of just discovered technique of
Gegenbauer polynomials in the position space
\cite{Chetyrkin:1980pr}):
\bea
L_0 &=& (q^2)^{-2-3\ep}\, 20\,\zeta_5 + {\cal O}(\ep), \ \ \ \   N_0 = (q^2)^{-2-3\ep}\, 20\,\zeta_5+ {\cal O}(\ep), 
\nonumber
\\
N_1 &=& (q^2)^{-1-3\ep}\, 20\,\zeta_5+ {\cal O}(\ep), \ \ \ \    N_2 = (q^2)^{-1-3\ep}\, 20\,\zeta_5+ {\cal O}(\ep),
\label{puzzle}
\eea
where $L_0,N_0,N_1$ and $N_2$  are  scalar three-loop p-integrals  (see Fig.~\ref{finite:3loop}).

\SetScale{1.2}
\SetWidth{.8}
\setlength{\unitlength}{1pt}

\begin{figure}[here]
\begin{center}
\thicklines
\parbox{50pt}{
\begin{picture}(50,60)(25,6) 
\put(50,0){$L_0$}
\CArc(46,30)(15,0,360)
\Line(37,42)(37,17)
\Line(55,42)(55,17)
\Line(30,30)(20,30)
\Line(62,30)(72,30)
\end{picture}
}
\hfill
$\bbuildrel{=\!=\!=}_{q^2=1}^{\ep=0}$
\hfill
\parbox{50pt}{
\begin{picture}(50,60)(27,6) 
\put(50,0){$N_0$}
\CArc(46,30)(15,0,360)
\Line(35,41)(57,19)
\Line(57,41)(48,32)
\Line(35,19)(43.4852813742386,27.4852813742386)
\Line(30,30)(20,30)
\Line(62,30)(72,30)
\end{picture}
}
\hfill
$\bbuildrel{=\!=\!=}_{q^2=1}^{\ep=0}$
\hfill
\parbox{50pt}{
\begin{picture}(54,60)(25,6) 
\put(50,0){$N_1$}
\CArc(46,30)(15,0,360)
\Line(46,45)(36,20)
\Line(46,45)(56,20)
\Line(30,30)(20,30)
\Line(62,30)(72,30)
\end{picture}
}
\hfill
$\bbuildrel{=\!=\!=}_{q^2=1}^{\ep=0}$
\hfill
\parbox{50pt}{
\begin{picture}(54,60)(28,6) 
\put(50,0){$N_2$}
\CArc(46,30)(15,0,360)
\Line(46,46)(46,14)
\Line(20,30)(46,30)
\Line(62,30)(72,30)
\end{picture}
}
\hfill
$= 20\,\zeta_5$
\hfill
\end{center}
\caption{Four  finite three-loop p-integrals  displaying a remarkable 
feature of being equal at $\ep =0 $ and $q^2=1$. }
\label{finite:3loop}
\end{figure}

Indeed, a short look  on eqs. \re{puzzle} immediately  leads to  an obvious question: 
why on the Earth four quite different (but all finite) p-integrals
have identical values at $D=4$ (if one set $q^2=1$)?
Incidentally, by that  time  a pioneering\footnote{To  our
knowledge it was the first full calculation of the $\beta$-function in
a four-dimensional model in {\bf four} loops.}  calculation of the
four-loop $\beta$-function in the $\phi^4$-model \cite{Kazakov:1979ik} had been just finished.
One of its  results was  the UV divergence of  the following  four-loop vertex-type integral 
\beq
\mathbf{UV}\left(
\parbox{2.23cm}{{\includegraphics[scale=0.75, bb=109 596 193 690]{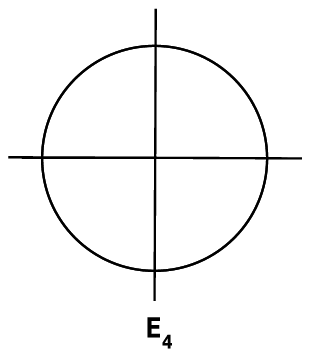}}}\right) = \frac{5\,\zeta_5}{\ep}.
\label{E4}
\eeq
The suspicious appearance of one and the same irrational constant
$\zeta_5$ with very simple coefficients in eqs. \re{puzzle} and
\re{E4} was suggesting some mysterious connection between three-loop {\em
finite} p-integrals $L_0, N_0,\, N_1, \, N_2$ and the {\em divergent}
part of the four-loop vertex-type integral $E_4$. In addition, a closer
inspection of all five diagrams revealed that the four propagators-type
diagram could be formally produced from the vertex graph in two steps (see Fig~\ref{glueing}):

\begin{figure}
\begin{center}
\includegraphics[height=0.20\textwidth]{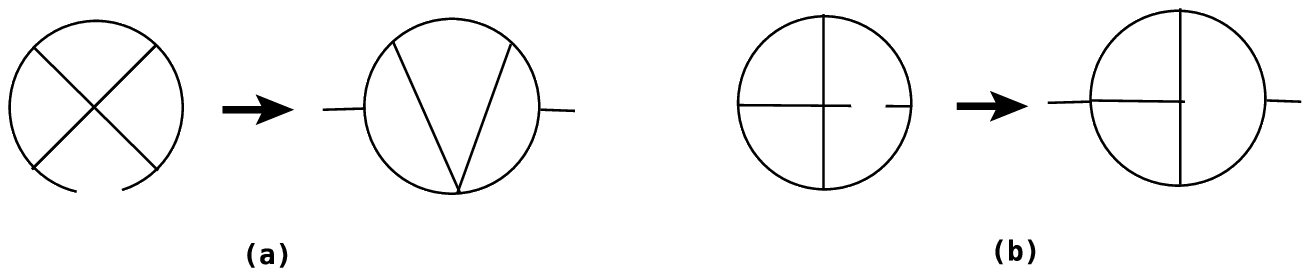}
\vspace{1cm}

\includegraphics[height=0.20\textwidth]{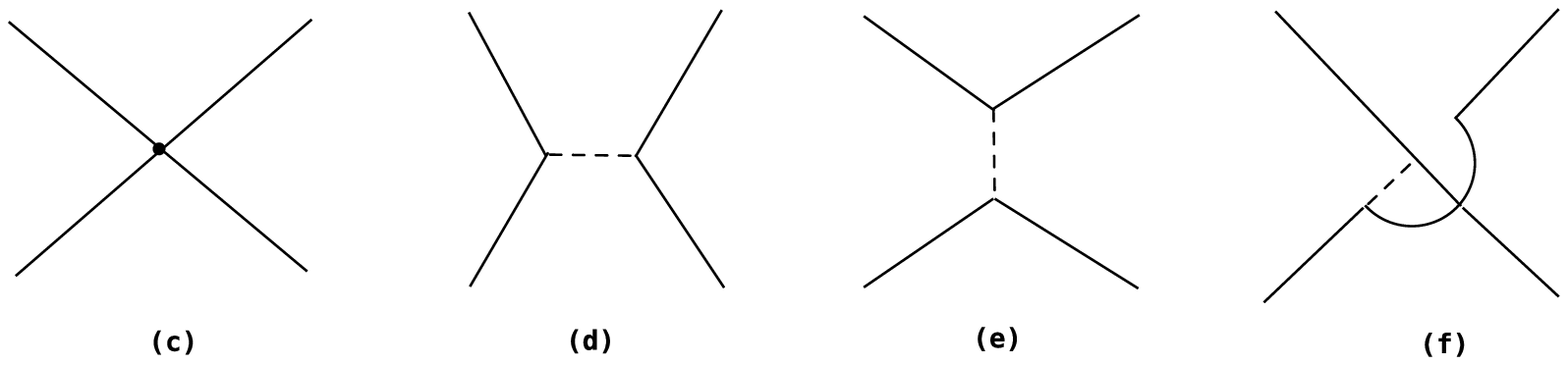}
\end{center}
\caption{\label{glueing} 
Two ways of cutting a line in the  graph $E_4$ (a,b).
The generic four-linear vertex (c) and three ways transforming it into 
a pair of the three-linear ones connected by an auxiliary propagator (d,e,f).
 }
\end{figure}

\noindent
(i)  Delete all four external lines from the
vertex diagram (transforming it, thus,  to a vacuum one).

\noindent
(ii) Cut in the resulting vacuum diagram either a line (there exist  only two non-equivalent choices, leading to $N_1$ and
$N_2$) or delete the central vertex\footnote{ The operation of deleting  of a
vertex means that one first transforms the vertex into two new ones by
introducing a fictitious line (with the unit propagator) and then
cutting the new line.  Note that deleting  a three-linear vertex does
not produce any new diagrams in addition to those coming from cutting
the corresponding three incident lines.  In general, one can cut a
four-linear vertex by a three non-equivalent ways a shown by
Fig~\ref{glueing}(d,e,f). Due to the high symmetry of the envelope
$E_4$ diagram the possibility $(d)$ and $(e)$ lead to one and the same
result.} (again, one could do it in two ways, leading to $N_0$ and
$L_0$).

The puzzle was finally understood after the geometrical construction was provided
with analytical content. As a result the Glue-and-Cut (GaC) symmetry of
massless propagators was established. Below we  prove a
theorem \cite{Chetyrkin:1981qh}
which solves  the puzzle. 

 Let $\langle \Ga \rangle (\ep)$ be a dimensionally regulated massless scalar
 $L+1$ loop vacuum Feynman amplitude {\em without} any subdivergences
 and with the superficial divergence index $\omega_\Ga = 0$ at
 four-dimensions. Surely, every expert would cry at this point that such an
 object is identical zero due to absence of any intrinsic scale which is true, 
 beyond any doubts\footnote{ By the way,
 a real mathematical justification why one could {\em
 self-consistently} set zero such massless vacuum integrals {\em
 within} dimensional regularization was to best of our knowledge
 performed only in \cite{Chetyrkin:1984wh}. It requires first of all a self-consistent
 and mathematically solid definitions of the very dimensional regularization  which, in turn,
 demands a heavy use of various parametric representations.}.
Please, be patient! In fact, by a Feynman amplitude we understand a
formal triplet consisting of the corresponding Feynman graph $\Ga$,
properly constructed Feynman {\em integrand} and, at last, a function
of kinematical parameters (external momenta and  masses)
resulting after evaluation the integrand. It means that even if the
function vanishes  for some particular choice of the
kinematical parameters it may become  nonzero after some modification of 
the latter.

Without essential loss of generality we assume that
the  graph $\Ga$ contains only 
triple
vertexes.  Consider an arbitrary line of $\ell$ of $\Ga$ with
\[
P_{\ell}(q) = \frac{{\cal P}(q,\dots)}{q^2}
\]
being the corresponding propagator of $\Ga$ and ${\cal P}(q)$ being
some polynomial in the line momentum~$q$. We allow the integral $ \langle \Ga \rangle $
to contain non-trivial numerator; in the case of the line $\ell$ being
a fictitious one with the unit propagator, one can always redefine 
 the  propagator  as follows:
\[
P_{\ell}(q) = \frac{{\cal P}'(q)}{q^2}  \ \ \mbox{with} \ \ 
{\cal P}'(q) =    q^2
{}.
\]
Let $\langle \Ga \rangle (m_0,\ep)$ be the Feynman amplitude obtained from   $\langle \Ga \rangle (\ep)$ by introducing 
an auxiliary non-zero mass $m_0$ into the selected propagator, as an infrared regulator, that is,
\[
P_{\ell}(q) \to P_{\ell}(q) \, \frac{q^2}{q^2+m_0^2} =  \frac{{\cal P}'(q,\dots)}{q^2+m_0^2}
{}.
\] 
Assuming that the loop momenta in $\langle \Ga\rangle $ are
chosen in such a way that
the momentum $q$ is the loop one, we can formally present the integral  $\langle \Ga\rangle (m_0,\ep)$ as a 
convolution
\beq
\langle \Ga \rangle   (m_0,\ep)
 = \int \langle \Ga \backslash \ell\rangle (q,\ep)\   \  \frac{\unl{d} q^D}{q^2+m_0^2}
\label{convolution}
{},
\eeq
where $\langle \Ga \backslash \ell\rangle (q,\ep)$ is the L-loop p-integral obtained from $\langle \Ga\rangle (m_0,\ep) $, first,
by ``freezing'' integration over $q$  
and, second, by multiplying the result by $(q^2+m_0^2)$.

\noindent
{\bf Theorem 3.} Under the above listed conditions the following statements are true

\begin{itemize}

\item[{\bf (a)}]
The vacuum integral $\langle \Ga\rangle( m_0,\ep)$ is IR  finite and  
its UV divergence is  a simple pole, that is  
\beq
\langle \Ga  \rangle (m_0,\ep) = \frac{C}{(L +1)\ep} + {\cal O}(\ep^0) 
{},
\label{T3:a}
\eeq
with $C$ being  a constant;

\item[{\bf (b)}] for every choice of the line $\ell$ the p-integral $ \langle \Ga \backslash \ell\rangle (q,\ep)  $ is
finite   and its value at $D=4$
meets  the condition
\beq 
\lim_{\ep\to  0} \langle \Ga \backslash \ell\rangle  (q,\ep) \,\, = \, \frac{C}{q^2}
\label{T3:b}
{}.
\eeq
\end{itemize}

\noindent
{\em \bf Proof.}

\begin{itemize}

\item[{\bf (a)}] Due to the assumed absence of  any subdivergences, the insertion of a mass does not influence  the 
UV divergence, but removes the only possibility for 
the IR one (related with integration over the region of 
{\em all} loop momenta
being  small). 
On dimensional grounds  we have
\beq
\langle \Ga \rangle (m_0,\ep) = (m_0^2)^{-(L+1)\ep}\, f(\ep)
\label{T3:proof:1}
{},
\eeq
with $f(\ep)$ depending  {\em only } on $\ep$.  Without any UV
subdivergences, the (minimal) UV counterterm corresponding to the
integral $\langle G\rangle (m_0,\ep)$ as a whole reduces to its pole part.  If
$f(\ep)$ would  contain a non-simple pole it would lead to appearance a
non-polynomial dependence on the mass $m_0$ of the counterterm in the 
direct violation of Theorem 1.  Thus,
\beq
C  = (L+1)\lim_{\ep \to 0 } \ep\, f(\ep)
\label{T3:proof:2}
{}.
\eeq

\item[{\bf (b)}]

The  (scalar) p-integral  $\langle \Ga \backslash \ell\rangle $ has a homogeneous dependence on $q^2$, namely
\beq
\langle \Ga \backslash \ell\rangle (q,\ep) =  (q^2)^{ -1 -L \ep}\, g(\ep)
\label{T3:proof:3}
\eeq
with $g(\ep)$ depending  {\em only } on $\ep$. As a consequence, the integral over 
$q$ in \re{convolution} can be  easily performed
with the help of a  textbook formula (see, e.g. \cite{Smirnov:2004ym}):
\begin{eqnarray*}
\lefteqn{\int \frac{\unl{d}^D \ell}{(m^2+\ell^2)\,(\ell^{2})^{1+n\,\ep}} =}\\
&{}&
( m^2)^{D/2-2-n\ep} \ \
\frac{\Gamma((n+1)\,\ep)\,\Gamma(1-(n+1)\,\ep)}{\Gamma(2-\ep)}
=
\frac{1}{(n+1)\ep} \cdot (1+{\cal O}(\ep))
\end{eqnarray*}
with the result
\beq
\langle \Ga \rangle (m_0,\ep) = \frac{g(\ep)}{(L+1) \,\ep} +  {\cal O}(\ep^0)   
\label{T3:proof:4}
{}.
\eeq
A comparison of eqs.~ (\ref{T3:proof:1},\ref{T3:proof:2})  and~\re{T3:proof:4} directly leads to 
eq.
\[
\lim_{\ep \to 0 }  g(\ep)  = C
{}
\]
which is equivalent to 
eq.~\re{T3:b}. $\Box$.

\end{itemize}

The GaC symmetry, proven in {\bf Theorem 3},  clearly explains the origin  of relations 
displayed in Fig.~\ref{finite:3loop}.  Still, considered by itself, it is not especially useful as
it does not provide us with the value of the  constant $C$.


\subsection{Three-loop master integrals from  glueing \label{3loop:gluieng2}   }

The situation is radically changed if one utilizes  the GaC symmetry 
{\em together}  with  the reduction to masters\footnote{Unfortunately, 
this step was overlooked thirty years ago, presumably, because the problem of evaluation of 
three-loop p-masters had been already solved {\em before} the idea of  glueing  appeared.}.
Indeed, let us forget for the  moment about eqs.~(\ref{puzzle}) and use {\em only} the GaC symmetry for
the  p-integrals shown on Fig.~\ref{finite:3loop} . This leads  to four equations, namely:
\beq 
N_0  = L_0 +  {\cal O}(\ep), \ \ N_0  = N_1 +  {\cal O}(\ep), \ \  \ N_0  = N_2 +  {\cal O}(\ep),
 \  \ N_0  =  {\cal O}(\ep^0)
\label{eqs_from_Gl:3lB}
{}.
\eeq
On the other hand the  reduction of three reducible p-integrals in eq.~\re{eqs_from_Gl:3lB} to masters gives:
\bea
L_0 &=&  
\frac{3\,(3D-10)\,(D-3)}{(D-4)^2}                  \,L_1
+\frac{4\,(D-3)^2}{(D-4)^2}                       \,P_4
+\,\frac{32\,(2D-7)\,(D-3)^2}{(D-4)^3}             \,P_1
\label{L0}
\\
&-&\,\frac{12(3D-8)\,(3D-10)\,(D-3)}{(D-4)^3}        \,P_2
 + \,\frac{4(9D^2-65 D+118)\,(3D-8)\,(2D-5)}{(D-4)^4} \,P_3
\nn
{},
\\
N_1 &=&  
\frac{(3D-10)\,(D-3)}{(D-4)^2}                    \,L_1  
+\,\frac{8(2D-7)\,(D-3)^2}{(D-4)^3}              \,P_1  
\label{N1}
\\
&-&\,\frac{2 (3D-8)\,(3D-10)\,(D-3)}{(D-4)^3}   \,P_2  
+\,\frac{4(3D-8)\,(2D-5)}{(D-4)^2}               \,P_3
\nonumber
{},
\\
N_2 &=&
-\,\frac{16(2D-7)\,(D-3)^2}{(D-4)^3} \,P_1             
+\frac{(3 D-8)\,(3D-10)\,(D-3)}{(D-4)^3} \,P_2             
\label{N2}
\\
&{+}&
\frac{10(3D-8)\,(3D-10)\,(2D-5)\,(2D-7)}{(D-4)^4}\,P_3
\nonumber
{}.
\eea

Now, it is well-known fact that a maximal order of the pole in $\ep$ of a (dimensionally regulated) L-loop p-integral
can not exceed\footnote{
In \cite{Chetyrkin:1984xa}
 the statement was proved for an arbitrary euclidean Feynman integral.} $L$.
Thus, we can  parametrize the coefficients  of the three-loop master p-integrals as follows\footnote{
Below the explicit result for the simplest master integral, $P_4$ is taken as granted; 
this fixes the global normalization of all remaining integrals.}:
\bea 
   N_0 &=& \sum_{i=-3}^0 N_{0,i}\,\ep^i + {\cal O}(\ep), 
\ \ L_1 = \sum_{i=-3}^2 L_{1,i}\,\ep^i+ {\cal O}(\ep^3),
\nonumber
\\
         P_1 &=& \sum_{i=-3}^3 P_{1i}\,\ep^i+ {\cal O}(\ep^4),
 \ \     P_2  =  \sum_{i=-3}^3 P_{2,i}\,\ep^i+ {\cal O}(\ep^4), 
\nonumber
\\
      P_3 &=&  \sum_{i=-3}^4 P_{3,i}\,\ep^i+ {\cal O}(\ep^5),
  \ \ P_4  =  \frac{P_{4,-3}}{\ep^3}, \ \   P_{4,-3} =1
{}.
\label{generic_expns:3l}
{}
\eea
Note that the higher term in $\ep$, not shown explicitly on
\re{generic_expns:3l}, at any case can not, obviously, be constrained
by eqs.~\re{generic_expns:3l}. In addition, they could contribute only to terms of
order $\ep$ or higher to the value of an arbitrary three-loop
p-integral (this statement follows from the knowledge of maximal power
of spurious pole in $\ep$ which might appear in front of a master
integral in the process of reduction, see
Fig.~\ref{3:loop:masters}).

After substitution of eqs.~\re{generic_expns:3l}, eqs.~\re{eqs_from_Gl:3lB} produce
some non-trivial constraints on coefficients of the $\ep$-expansion of
our master p-integrals.
To be specific, let consider first eq. from \re{eqs_from_Gl:3lB}. 
Its expanded in $\ep$ form reads
\beq
\lim_{\ep \to 0} \ep^n (L_0 - N_0) =0 
\label{eqL0}
\eeq
for $ n \ge 0$.  Note that \re{eqL0} is met identically if  $  n >  7$  
(because of the fact that the maximal allowed   poles
in $\ep$ which could appear in eqs.~(\ref{L0},\ref{N1},\ref{N2}) and eqs~(\ref{generic_expns:3l}), are 4 and 3 respectively).
For $n=7,6,5$ and $4$ the resulting equations are\footnote{Note, that for brevity in writing  eqs.~(\ref{n=6}-\ref{n=4}) 
we have used eq.~\ref{n=7}
to discard the  terms proportional to (zero) coefficient $P_{{3,-3}}$.}
\bea
0&=& 
\,P_{{3,-3}}  
\label{n=7},
\\
0&=& 
6 \,P_{{3,-2}}+12 \,P_{{2,-3}}-4 \,P_{{1,-3}}
\label{n=6},
\\
0&=&
-59 \,P_{{3,-2}}+6 \,P_{{3,-1}}-78%
 \,P_{{2,-3}}+12 \,P_{{2,-2}}+32 \,P_{{1,-3}}-4%
 \,P_{{1,-2}}+\frac{3 L_{{1,-3}}}{2}+1
\label{n=5}
{},
\\
0&=&
239 \,P_{{3,-2}}-59 \,P_{{3,-1}}+6%
 \,P_{{3,0}}+162 \,P_{{2,-3}}-78 \,P_{{2,-2}}+12 \,P_{{2,-1}}
\\
&{}&-80 \,P_{{1,-3}}+32 \,P_{{1,-2}}-4 \,P_{{1,-1}}-\frac{15%
 L_{{1,-3}}}{2}+\frac{3 L_{{1,-2}}}{2}-4 
\label{n=4}
{}.
\eea
Already now we can see that eqs.~(\ref{n=5}) and \re{n=4} express two coefficients,  $L_{{1,-3}}$ and $L_{{1,-2}}$,
of a non-primitive   integral through  the coefficients of primitive ones, namely, $P_1$, $P_2$ and $P_3$.
Indeed, a  solution of eqs.~(\ref{n=7}-\ref{n=4}) is
\bea
P_{\text{3,-3}}&=&0 \label{epm7},
 \\ 
P_{\text{1,-3}}&=&\frac{3 P_{\text{3,-2}}}{2}+3 P_{\text{2,-3}} 
\label{epm6}
,
  \\ 
L_{\text{1,-3}}&=&\frac{22 P_{\text{3,-2}}}{3}-4 P_{\text{3,-1}}-12
  P_{\text{2,-3}}-8 P_{\text{2,-2}}+\frac{8 P_{\text{1,-2}}}{3}-\frac{2}{3}\
  \label{epm5} 
,
\\ 
L_{\text{1,-2}}&=&-\frac{128 P_{\text{3,-2}}}{3}+\frac{58 P_{\text{3,-1}}}{3}
  -4 P_{\text{3,0}}-8 P_{\text{2,-3}}+12 P_{\text{2,-2}}-8\
  P_{\text{2,-1}}-8 P_{\text{1,-2}}+\frac{8 P_{\text{1,-1}}}{3}-\frac{2}{3}\
  \label{epm4}  
{}.
\eea
Proceeding in the same vein we arrive eventually  to a linear system of  31  equations (not necessarily independent)
for 32  coefficients 
$N_{0,i_0}, L_{1,i_1},  P_{1,j_1}, P_{2,j_2}, P_{3,j_3}$. 
One can solve the system by expressing the coefficients from  more complicated masters  through  those from  less complicated ones.
A  convenient   ordering is given by two rules\footnote{The order between primitive integrals  is rather arbitrary, except for 
the natural choice to use $P_4$ as the easiest one.}
\begin{itemize}
\item
 $N_{0,i_0} \succ   L_{1,i_1}  \succ P_{1,j_1}  \succ P_{2,j_2}  \succ P_{3,j_3}  \succ P_{4,j_4}$.
\item For two   coefficients of a  master integral
 the more complicated one is  that 
 with larger   value of the second index.
\end{itemize}
The system is easily solved with the  result:  coefficients 
are expressed in terms of only   {\rm eight } coefficients of three {\em primitive} 
integrals, namely,
\beq
P_{3,-1,}, P_{3,0}, P_{3,1}, P_{3,2}, P_{3,3}, P_{3,4},  P_{2,3} ,  P_{4,-3}
{}.
\label{input_coef:3l}
\eeq
These  eight coefficients are trivially determined from eqs.~\re{P1_P4} and,
finally, we arrive at the following results for the   three-loop  master integrals:
\newcommand{\boxeD}[1]{\hspace{1.mm}\boxed{#1}\hspace{1.mm}}

\begin{align}
\begin{minipage}{2.7cm}{
\begin{center}
\begin{picture}(50,80)(15,-4.0)
\put(15,0){$N_0,\ep^0$}
\CArc(26,30)(15,0,360)
\Line(15,41)(37,19)
\Line(37,41)(28,32)
\Line(15,19)(23.4852813742386,27.4852813742386)
\Line(10,30)(5,30)
\Line(42,30)(47,30)
\end{picture}
\end{center}
}
\end{minipage}
{}&
\hspace{-1.0mm}
=20 \,\zeta_{5} 
+ \,{\cal O}(\varepsilon)
\label{N0res},
\\ 
\displaybreak[1]
\begin{minipage}{2.7cm}{
\begin{center}
\begin{picture}(50,80)(15,-4.0)
\put(20,0){$L_1,\ep^2$}
\CArc(25,30)(15,0,360)
\CArc(35,30)(15,0,360)
\Line(9,30)(4,30)
\Line(51,30)(56,30)
\end{picture}
\end{center}
}
\end{minipage}
{}&
\hspace{-1.0mm}
=\frac{ 1 }{3 \,\ep^3}+\frac{ 1 }{3 \,\ep^2}+\frac{ 1 }{3 \,\ep }-\frac{7}{3}+\frac{14 \,\zeta_{3} }{3}
+ \,\ep \left( -\frac{67}{3}+\frac{14 \,\zeta_{3} }{3}
+7 \,\zeta_{4} 
\right)\nonumber \displaybreak[1] \\ & 
+ \,\ep^2\left( -\frac{403}{3}+\frac{86 \,\zeta_{3} }{3}
+7 \,\zeta_{4} 
+126 \,\zeta_{5} 
\right)+\,{\cal O}(\varepsilon^3)
\label{L1},
\\ 
\displaybreak[1]
\begin{minipage}{2.7cm}{
\begin{center}
\begin{picture}(50,80)(15,-4.0)
\put(15,0){$P_1,\ep^3$}
\CArc(26,30)(15,0,360)
%
\Line(26,45.1)(42,30)
\Line(26,45.1)(10,30)
\Line(10,30)(5,30)
\Line(42,30)(47,30)
\end{picture}
\end{center}
}
\end{minipage}
{}&
\hspace{-1.0mm}
=-\frac{ 1 }{3 \,\ep^2}-\frac{4}{3 \,\ep }-\frac{16}{3}+ \,\ep \left( -\frac{64}{3}+\frac{16 \,\zeta_{3} }{3}
\right)+ \,\ep^2\left( -\frac{256}{3}+\frac{64 \,\zeta_{3} }{3}
+8 \,\zeta_{4} 
\right)\nonumber \displaybreak[1] \\ & 
+ \,\ep^3\left( -\frac{1024}{3}+\frac{256 \,\zeta_{3} }{3}
+32 \,\zeta_{4} 
+64 \,\zeta_{5} 
\right)+\,{\cal O}(\varepsilon^4)
\label{P1},
\\ 
\displaybreak[1]
\begin{minipage}{2.7cm}{
\begin{center}
\begin{picture}(50,80)(15,-4.0)
\put(15,0){$P_2,\ep^3$}
\CArc(10,30)(10,0,360)
\CArc(36,30)(15,0,360)
\Line(0,30)(-5,30)
\Line(20,30)(57,30)
\end{picture}
\end{center}
}
\end{minipage}
{}&
\hspace{-1.0mm}
=-\frac{ 1 }{4 \,\ep^2}-\frac{5}{8 \,\ep }-\frac{27}{16}+ \,\ep \left( -\frac{153}{32}+\frac{3 \,\zeta_{3} }{2}
\right)+ \,\ep^2\left( -\frac{891}{64}+\frac{15 \,\zeta_{3} }{4}
+\frac{9 \,\zeta_{4} }{4}
\right)\nonumber \displaybreak[1] \\ & 
+ \,\ep^3
\left(
\boxeD{
 -\frac{\normalsize 5265}{128}
+\frac{81 \,\zeta_{3} }{8}
+\frac{45 \,\zeta_{4} }{8}
+\frac{21 \,\zeta_{5} }{2}
}
\right)
+\,{\cal O}(\varepsilon^4)
\label{P2},
\\ 
\displaybreak[1]
\begin{minipage}{2.7cm}{
\begin{center}
\begin{picture}(50,80)(15,-4.0)
\put(10,0){$P_3,\ep^4$}
\CArc(20,34)(11,0,360)
\CArc(20,26)(11,0,360)
\Line(9,30)(4,30)
\Line(31,30)(36,30)
\end{picture}
\end{center}
}
\end{minipage}
{}&
\hspace{-1.0mm}
=\boxeD{\frac{ 1 }{36}}\frac{1}{ \,\ep }+\boxeD{\frac{35}{216}} 
+
\boxeD{
\frac{991}{1296}
}
\,\varepsilon
+ \,
\ep^2\left(
\boxeD{ 
\frac{26207}{7776}-\frac{11 \,\zeta_{3} }{18}
}
\right)
\\ & 
+ \,\ep^3\left(
\boxeD{
 \frac{670951}{46656}-\frac{385 \,\zeta_{3} }{108}
-\frac{11 \,\zeta_{4} }{12}
}
\right)\nonumber \displaybreak[1] \\ & 
+ \,\ep^4\left(
\boxeD{
 \frac{16852031}{279936}-\frac{10901 \,\zeta_{3} }{648}
-\frac{385 \,\zeta_{4} }{72}
-\frac{13 \,\zeta_{5} }{2}
}
\right)+\,{\cal O}(\varepsilon^5)
\label{P3},
\\ 
\displaybreak[1]
\begin{minipage}{2.7cm}{
\begin{center}
\begin{picture}(50,80)(15,-4.0)
\put(29,0){$P_4,\ep^2$}
\CArc(18,30)(7.5,0,360)
\CArc(33,30)(7.5,0,360)
\CArc(48,30)(7.5,0,360)
\Line(10,30)(5,30)
\Line(56,30)(61,30)
\end{picture}
\end{center}
}
\end{minipage}
{}&
=
\hspace{2.0mm}
 \frac{\boxeD{1}}{\ep^3} +\,{\cal O}(\varepsilon^3)
\label{P4}
{},
\end{align}

\noindent
where we have boxed  the  eight  input coefficients. The comparison with eqs.~(\ref{P1_P4},\ref{puzzle}) and \re{L0}
demonstrates the all unboxed coefficients  have  been correctly determined
through the gluing  procedure.  

A remarkable feature of the above discussed, glue-and-cut based,
determination of the three-loop masters is that the both non-primitive
(read non-trivial) master integrals $N_0$ and $L_1$ have been
expressed through essentially trivial (read primitive) FI's. Even more,
as many as eleven coefficients of the primitive MI's $P_1$ and $P_2$
(see eqs.~(\ref{P1},\ref{P2})) have also been fixed through only  eight 
coefficients listed in \re{input_coef:3l}. Thus, we see that integration by parts  
identities  together with the glue-and-cut symmetry severely constrain the  
{\em values } of master integrals. 


\section{Four-loop Integrals \label{4loop}}

\subsection{Four-loop master integrals from glueing \label{sol4}}

Following the same procedure in the case of four-loop propagator massless integrals,
one should consider all possible cuttings of a set of   five-loop vacuum massless 
diagrams  with integrand of mass dimension  {\em twenty} and without subdivergences
(or, equivalently, superficially {and} logarithmically divergent).  

Again, as in three-loop case, GaC  relations provide us with enough information to 
express all the  necessary  coefficients of the $\ep$-expansions of all MI's  through 
some trivial integrals.
The number of the input five-loop tadpoles 
\ice{(1044)}
and the resulting relations \ice{($044\cdot12 = 12528$)} (around a hundred and a thousand respectively)
 are too large to be presented here, so in the equations
to  follow we give only  the final results.

\SetScale{1.0}
\begin{align}
{}&
\begin{minipage}{2.7cm}{
\begin{center}
\begin{picture}(50,80)(35,10)
\CArc(50,50)(20,0,360)
\Line(70,50)(80,50)
\Line(30,50)(20,50)
\Line(61,67)(61,33)
\Line(39,67)(39,34)
\Line(39,50)(61,50)
\put(33,14){$M_{61},\,\ep^1$}
\end{picture}
\end{center}
}
\end{minipage}
\hspace{-5mm}
=-\frac{10 \sbz \zeta_{5}}{ \,\ep }+50 \sbz \zeta_{5}
-10 \,\zeta_{3}^2 
-25 \zeta_6 
\nonumber \displaybreak[1] \\ & 
+ \,\ep \left( 90 \sbz \zeta_{5}
+50 \,\zeta_{3}^2 
+125 \zeta_6 
-30 \,\zeta_{3}\, \zeta_4 
+\frac{19 \,\zeta_{7}}{2}
\right)+\,{\cal O}(\varepsilon^2)
\label{m61},
\\ 
\displaybreak[1]
{}&
\begin{minipage}{2.7cm}{
\begin{center}
\begin{picture}(50,80)(35,10)
\CArc(50,50)(20,0,360)
\Line(70,50)(80,50)
\Line(30,50)(20,50)
\Line(62,34)(62,47)
\Line(38,34)(38,47)
\Line(38,47)(62,47)
\Line(62,47)(36,64.3333333333333)
\Line(38,47)(46.3205029433784,52.5470019622523)
\Line(64,64)(53,56.6666666666667)
\put(33,14){$M_{62},\,\ep^0$}
\end{picture}
\end{center}
}
\end{minipage}
\hspace{-5mm}
=-\frac{10 \sbz \zeta_{5}}{ \,\ep }+130 \sbz \zeta_{5}
-10 \,\zeta_{3}^2 
-25 \zeta_6 
-70 \,\zeta_{7}
+ \,{\cal O}(\varepsilon)
\label{m62},
\\ 
\displaybreak[1]
{}&
\begin{minipage}{2.7cm}{
\begin{center}
\begin{picture}(50,80)(35,10)
\CArc(50,50)(20,0,360)
\Line(70,50)(80,50)
\Line(30,50)(20,50)
\Line(39,67)(39,33)
\Line(39,50)(68,57.25)
\Line(61,33)(61,53)
\Line(61,67)(61,58)
\put(33,14){$M_{63},\,\ep^0$}
\end{picture}
\end{center}
}
\end{minipage}
\hspace{-5mm}
=-\frac{5 \sbz \zeta_{5}}{ \,\ep }+45 \sbz \zeta_{5}
-41 \,\zeta_{3}^2 
-\frac{25 \zeta_6 }{2}
+\frac{161 \,\zeta_{7}}{2}
+ \,{\cal O}(\varepsilon)
\label{m63},
\\ 
\displaybreak[1]
{}&
\begin{minipage}{2.7cm}{
\begin{center}
\begin{picture}(50,80)(35,10)
\CArc(50,50)(20,0,360)
\Line(70,50)(80,50)
\Line(30,50)(20,50)
\Line(50,70)(34,38)
\Line(40,50)(65,37.5)
\Line(50,70)(50,48)
\Line(50,30)(50,42)
\put(33,14){$M_{51},\,\ep^1$}
\end{picture}
\end{center}
}
\end{minipage}
\hspace{-5mm}
=-\frac{5 \sbz \zeta_{5}}{ \,\ep }+45 \sbz \zeta_{5}
-17 \,\zeta_{3}^2 
-\frac{25 \zeta_6 }{2}
\nonumber \displaybreak[1] \\ & 
+ \,\ep \left( -195 \sbz \zeta_{5}
+153 \,\zeta_{3}^2 
+\frac{225 \zeta_6 }{2}
-51 \,\zeta_{3}\, \zeta_4 
-\frac{85 \,\zeta_{7}}{2}
\right)+\,{\cal O}(\varepsilon^2)
\label{m51},
\\ 
\displaybreak[1]
{}&
\begin{minipage}{2.7cm}{
\begin{center}
\begin{picture}(50,80)(35,10)
\CArc(50,50)(20,0,360)
\Line(61,67)(52,53.5)
\Line(39,34)(48,47.5)
\Line(50,50)(39,66.5)
\Line(50,50)(61,33.5)
\Line(61,67)(39,67)
\Line(70,48)(80,48)
\Line(30,48)(20,48)
\put(33,14){$M_{41},\,\ep^1$}
\end{picture}
\end{center}
}
\end{minipage}
\hspace{-5mm}
=\frac{20 \sbz \zeta_{5}}{ \,\ep }-80 \sbz \zeta_{5}
-22 \,\zeta_{3}^2 
+50 \zeta_6 
\nonumber \displaybreak[1] \\ & 
+ \,\ep \left( 80 \sbz \zeta_{5}
+88 \,\zeta_{3}^2 
-200 \zeta_6 
-66 \,\zeta_{3}\, \zeta_4 
+\frac{4685 \,\zeta_{7}}{8}
\right)+\,{\cal O}(\varepsilon^2)
\label{m41},
\\ 
\displaybreak[1]
{}&
\begin{minipage}{2.7cm}{
\begin{center}
\begin{picture}(50,80)(35,10)
\CArc(50,50)(20,0,360)
\Line(61,66)(52,52.5)
\Line(39,34)(48,47.5)
\Line(50,50)(39,66.5)
\Line(50,50)(61,33.5)
\Line(61,66)(70,48)
\Line(70,48)(80,48)
\Line(30,48)(20,48)
\put(33,14){$M_{42},\,\ep^1$}
\end{picture}
\end{center}
}
\end{minipage}
\hspace{-5mm}
=\frac{20 \sbz \zeta_{5}}{ \,\ep }-80 \sbz \zeta_{5}
+8 \,\zeta_{3}^2 
+50 \zeta_6 
\nonumber \displaybreak[1] \\ & 
+ \,\ep \left( 80 \sbz \zeta_{5}
-32 \,\zeta_{3}^2 
-200 \zeta_6 
+24 \,\zeta_{3}\, \zeta_4 
+520 \,\zeta_{7}
\right)+\,{\cal O}(\varepsilon^2)
\label{m42},
\\ 
\displaybreak[1]
{}&
\begin{minipage}{2.7cm}{
\begin{center}
\begin{picture}(50,80)(35,10)
\CArc(50,50)(20,0,360)
\Line(50,50)(61,66.5)
\Line(50,50)(39,33.5)
\Line(70,48)(80,48)
\Line(30,48)(20,48)
\Line(61,67)(61,33)
\Line(39,67)(39,34)
\put(33,14){$M_{44},\,\ep^0$}
\end{picture}
\end{center}
}
\end{minipage}
\hspace{-5mm}
=\frac{441 \,\zeta_{7}}{8}+ \,{\cal O}(\varepsilon)
\label{m44},
\\ 
\displaybreak[1]
{}&
\begin{minipage}{2.7cm}{
\begin{center}
\begin{picture}(50,80)(35,10)
\CArc(50,50)(20,0,360)
\Line(61,66)(52,52.5)
\Line(39,34)(48,47.5)
\Line(50,50)(39,66.5)
\Line(50,50)(61,33.5)
\Line(61,67)(61,34)
\Line(70,48)(80,48)
\Line(30,48)(20,48)
\put(33,14){$M_{45},\,\ep^1$}
\end{picture}
\end{center}
}
\end{minipage}
\hspace{-5mm}
=36 \,\zeta_{3}^2 + \,\ep \left( 108 \,\zeta_{3}\, \zeta_4 
-378 \,\zeta_{7}
\right)+\,{\cal O}(\varepsilon^2)
\label{m45},
\\ 
\displaybreak[1]
{}&
\begin{minipage}{2.7cm}{
\begin{center}
\begin{picture}(50,80)(35,10)
\Line(30,50)(20,50)
\Line(70,50)(80,50)
\CArc(50,50)(20,0,360)
\Line(34,62)(66,62)
\Line(50,30)(34,62)
\Line(50,30)(66,62)
\put(33,14){$M_{34},\,\ep^3$}
\end{picture}
\end{center}
}
\end{minipage}
\hspace{-5mm}
=\frac{ 1 }{12\,\varepsilon^{4}}+\frac{ 1 }{4\,\varepsilon^{3}}+\frac{7}{12\,\varepsilon^{2}}+\frac{1}{\,\varepsilon}\left( -\frac{17}{12}
+\frac{25 \sbz \zeta_{3}}{6}
\right)-\frac{377}{12}
+\frac{25 \sbz \zeta_{3}}{2}
+\frac{25 \zeta_4 }{4}
\nonumber \displaybreak[1] \\ & 
+ \,\ep \left( -\frac{3401}{12}
+\frac{463 \sbz \zeta_{3}}{6}
+\frac{75 \zeta_4 }{4}
+\frac{465 \sbz \zeta_{5}}{2}
\right)\nonumber \displaybreak[1] \\ & 
+\,\varepsilon^{2}\left( -\frac{24497}{12}
+\frac{3031 \sbz \zeta_{3}}{6}
+\frac{463 \zeta_4 }{4}
+\frac{1395 \sbz \zeta_{5}}{2}
-\frac{1247 \,\zeta_{3}^2 }{6}
+\frac{3425 \zeta_6 }{6}
\right)\nonumber \displaybreak[1] \\ & 
+\,\varepsilon^{3}\left( -\frac{158273}{12}
+\frac{19663 \sbz \zeta_{3}}{6}
+\frac{3031 \zeta_4 }{4}
+\frac{6807 \sbz \zeta_{5}}{2}
  \right. \nonumber \\   & \hspace{1.4cm} \left.  
 -\frac{1247 \,\zeta_{3}^2 }{2}
+\frac{3425 \zeta_6 }{2}
-\frac{1247 \,\zeta_{3}\, \zeta_4 }{2}
+\frac{12503 \,\zeta_{7}}{2}
\right)+\,{\cal O}(\varepsilon^4)
\label{m34},
\\ 
\displaybreak[1]
{}&
\begin{minipage}{2.7cm}{
\begin{center}
\begin{picture}(50,80)(35,10)
\Line(30,50)(20,50)
\Line(50,50)(80,50)
\CArc(50,50)(20,0,360)
\Line(50,50)(41,68)
\Line(50,50)(41,32)
\Line(41,32)(41,68)
\put(33,14){$M_{35},\,\ep^2$}
\end{picture}
\end{center}
}
\end{minipage}
\hspace{-5mm}
=\frac{ \sbz \zeta_{3}}{2\,\varepsilon^{2}}+\frac{1}{\,\varepsilon}\left( \frac{3 \sbz \zeta_{3}}{2}
+\frac{3 \zeta_4 }{4}
\right)+\frac{19 \sbz \zeta_{3}}{2}
+\frac{9 \zeta_4 }{4}
-\frac{23 \sbz \zeta_{5}}{2}
\nonumber \displaybreak[1] \\ & 
+ \,\ep \left( \frac{103 \sbz \zeta_{3}}{2}
+\frac{57 \zeta_4 }{4}
-\frac{69 \sbz \zeta_{5}}{2}
+\frac{29 \,\zeta_{3}^2 }{2}
-30 \zeta_6 
\right)\nonumber \displaybreak[1] \\ & 
+\,\varepsilon^{2}\left( \frac{547 \sbz \zeta_{3}}{2}
+\frac{309 \zeta_4 }{4}
-\frac{437 \sbz \zeta_{5}}{2}
+\frac{87 \,\zeta_{3}^2 }{2}
-90 \zeta_6 
+\frac{87 \,\zeta_{3}\, \zeta_4 }{2}
-\frac{1105 \,\zeta_{7}}{4}
\right)+\,{\cal O}(\varepsilon^3)
\label{m35},
\\ 
\displaybreak[1]
{}&
\begin{minipage}{2.7cm}{
\begin{center}
\begin{picture}(50,80)(35,10)
\Line(20,50)(80,50)
\CArc(50,50)(20,0,360)
\Line(50,30)(50,70)
\put(33,14){$M_{36},\,\ep^1$}
\end{picture}
\end{center}
}
\end{minipage}
\hspace{-5mm}
=\frac{5 \sbz \zeta_{5}}{ \,\ep }-5 \sbz \zeta_{5}
-7 \,\zeta_{3}^2 
+\frac{25 \zeta_6 }{2}
+ \,\ep \left( 35 \sbz \zeta_{5}
+7 \,\zeta_{3}^2 
-\frac{25 \zeta_6 }{2}
-21 \,\zeta_{3}\, \zeta_4 
+\frac{127 \,\zeta_{7}}{2}
\right)+\,{\cal O}(\varepsilon^2)
\label{m36},
\\ 
\displaybreak[1]
{}&
\begin{minipage}{2.7cm}{
\begin{center}
\begin{picture}(50,80)(35,10)
\CArc(40,50)(10,0,360)
\CArc(66,50)(15,0,360)
\Line(55,61)(77,39)
\Line(77,61)(67.54,51.54)
\Line(64.46,48.46)(55,39)
\Line(30,50)(20,50)
\Line(82,50)(92,50)
\put(33,14){$M_{52},\,\ep^1$}
\end{picture}
\end{center}
}
\end{minipage}
\hspace{-5mm}
\hspace{7mm}=\frac{20 \sbz \zeta_{5}}{ \,\ep }-80 \sbz \zeta_{5}
+68 \,\zeta_{3}^2 
+50 \zeta_6 
\nonumber \displaybreak[1] \\ & 
+ \,\ep \left( 80 \sbz \zeta_{5}
-272 \,\zeta_{3}^2 
-200 \zeta_6 
+204 \,\zeta_{3}\, \zeta_4 
+450 \,\zeta_{7}
\right)+\,{\cal O}(\varepsilon^2)
\label{m52},
\\ 
\displaybreak[1]
{}&
\begin{minipage}{2.7cm}{
\begin{center}
\begin{picture}(50,80)(35,10)
\CArc(50,50)(20,0,360)
\Line(20,50)(80,50)
\Line(42,68)(60,50)
\Line(58,68)(51.6496,61.6496)
\Line(47.56,57.56)(40,50)
\put(33,14){$M_{43},\,\ep^1$}
\end{picture}
\end{center}
}
\end{minipage}
\hspace{-5mm}
=-\frac{5 \sbz \zeta_{5}}{ \,\ep }+45 \sbz \zeta_{5}
-17 \,\zeta_{3}^2 
-\frac{25 \zeta_6 }{2}
\nonumber \displaybreak[1] \\ & 
+ \,\ep \left( -195 \sbz \zeta_{5}
+153 \,\zeta_{3}^2 
+\frac{225 \zeta_6 }{2}
-51 \,\zeta_{3}\, \zeta_4 
-\frac{225 \,\zeta_{7}}{2}
\right)+\,{\cal O}(\varepsilon^2)
\label{m43},
\\ 
\displaybreak[1]
{}&
\begin{minipage}{2.7cm}{
\begin{center}
\begin{picture}(50,80)(35,10)
\CArc(19,50)(10,0,360)
\CArc(45,50)(15,0,360)
\CArc(55,50)(15,0,360)
\Line(9,50)(-1,50)
\Line(71,50)(81,50)
\put(33,14){$M_{32},\,\ep^3$}
\end{picture}
\end{center}
}
\end{minipage}
\hspace{-5mm}
=\frac{ 1 }{3\,\varepsilon^{4}}+\frac{ 1 }{3\,\varepsilon^{3}}+\frac{ 1 }{3\,\varepsilon^{2}}+\frac{1}{\,\varepsilon}\left( -\frac{7}{3}
+\frac{14 \sbz \zeta_{3}}{3}
\right)-\frac{67}{3}
+\frac{14 \sbz \zeta_{3}}{3}
+7 \zeta_4 
\nonumber \displaybreak[1] \\ & 
+ \,\ep \left( -\frac{403}{3}
+\frac{86 \sbz \zeta_{3}}{3}
+7 \zeta_4 
+126 \sbz \zeta_{5}
\right)\nonumber \displaybreak[1] \\ & 
+\,\varepsilon^{2}\left( -\frac{2071}{3}
+\frac{478 \sbz \zeta_{3}}{3}
+43 \zeta_4 
+126 \sbz \zeta_{5}
-\frac{226 \,\zeta_{3}^2 }{3}
+\frac{910 \zeta_6 }{3}
\right)\nonumber \displaybreak[1] \\ & 
+\,\varepsilon^{3}\left( -\frac{9823}{3}
+\frac{2446 \sbz \zeta_{3}}{3}
+239 \zeta_4 
+534 \sbz \zeta_{5}
-\frac{226 \,\zeta_{3}^2 }{3}
+\frac{910 \zeta_6 }{3}
-226 \,\zeta_{3}\, \zeta_4 
+1960 \,\zeta_{7}
\right)+\,{\cal O}(\varepsilon^4)
\label{m32},
\\ 
\displaybreak[1]
{}&
\begin{minipage}{2.7cm}{
\begin{center}
\begin{picture}(50,80)(35,10)
\CArc(50,50)(20,0,360)
\CArc(50,60)(10,0,360)
\CArc(50,40)(10,0,360)
\Line(30,50)(20,50)
\Line(70,50)(80,50)
\put(33,14){$M_{33},\,\ep^3$}
\end{picture}
\end{center}
}
\end{minipage}
\hspace{-5mm}
=\frac{ 1 }{6\,\varepsilon^{4}}+\frac{ 1 }{3\,\varepsilon^{3}}+\frac{ 1 }{3\,\varepsilon^{2}}+\frac{1}{\,\varepsilon}\left( -\frac{17}{3}
+\frac{31 \sbz \zeta_{3}}{3}
\right)-\frac{197}{3}
+\frac{62 \sbz \zeta_{3}}{3}
+\frac{31 \zeta_4 }{2}
\nonumber \displaybreak[1] \\ & 
+ \,\ep \left( -\frac{1529}{3}
+\frac{386 \sbz \zeta_{3}}{3}
+31 \zeta_4 
+449 \sbz \zeta_{5}
\right)\nonumber \displaybreak[1] \\ & 
+\,\varepsilon^{2}\left( -\frac{10205}{3}
+\frac{2510 \sbz \zeta_{3}}{3}
+193 \zeta_4 
+898 \sbz \zeta_{5}
-\frac{983 \,\zeta_{3}^2 }{3}
+\frac{3290 \zeta_6 }{3}
\right)\nonumber \displaybreak[1] \\ & 
+\,\varepsilon^{3}\left( -\frac{62801}{3}
+\frac{15974 \sbz \zeta_{3}}{3}
+1255 \zeta_4 
+4354 \sbz \zeta_{5}
  \right. \nonumber \\   & \hspace{1.4cm} \left.  
 -\frac{1966 \,\zeta_{3}^2 }{3}
+\frac{6580 \zeta_6 }{3}
-983 \,\zeta_{3}\, \zeta_4 
+11338 \,\zeta_{7}
\right)+\,{\cal O}(\varepsilon^4)
\label{m33},
\\ 
\displaybreak[1]
{}&
\begin{minipage}{2.7cm}{
\begin{center}
\begin{picture}(50,80)(35,10)
\CArc(40,50)(20,0,360)
\CArc(60,50)(20,0,360)
\Line(50,33)(20,48)
\Line(20,48)(10,48)
\Line(80,48)(90,48)
\put(33,14){$M_{21},\,\ep^4$}
\end{picture}
\end{center}
}
\end{minipage}
\hspace{-5mm}
\hspace{7mm}=-\frac{5}{48\,\varepsilon^{3}}-\frac{31}{96\,\varepsilon^{2}}-\frac{95}{192 \,\ep }+\frac{1133}{384}
-\frac{19 \sbz \zeta_{3}}{12}
\nonumber \displaybreak[1] \\ & 
+ \,\ep \left( \frac{30097}{768}
-\frac{233 \sbz \zeta_{3}}{24}
-\frac{19 \zeta_4 }{8}
\right)+\,\varepsilon^{2}\left( \frac{463349}{1536}
-\frac{3385 \sbz \zeta_{3}}{48}
-\frac{233 \zeta_4 }{16}
-\frac{341 \sbz \zeta_{5}}{4}
\right)\nonumber \displaybreak[1] \\ & 
+\,\varepsilon^{3}\left( \frac{6004105}{3072}
-\frac{46469 \sbz \zeta_{3}}{96}
-\frac{3385 \zeta_4 }{32}
-\frac{3187 \sbz \zeta_{5}}{8}
+\frac{493 \,\zeta_{3}^2 }{6}
-\frac{1255 \zeta_6 }{6}
\right)\nonumber \displaybreak[1] \\ & 
+\,\varepsilon^{4}\left( \frac{71426093}{6144}
-\frac{590281 \sbz \zeta_{3}}{192}
-\frac{46469 \zeta_4 }{64}
-\frac{33875 \sbz \zeta_{5}}{16}
  \right. \nonumber \\   & \hspace{1.4cm} \left.  
 +\frac{4673 \,\zeta_{3}^2 }{12}
-\frac{2915 \zeta_6 }{3}
+\frac{493 \,\zeta_{3}\, \zeta_4 }{2}
-\frac{16619 \,\zeta_{7}}{8}
\right)+\,{\cal O}(\varepsilon^5)
\label{m21},
\\ 
\displaybreak[1]
{}&
\begin{minipage}{2.7cm}{
\begin{center}
\begin{picture}(50,80)(35,10)
\CArc(50,50)(20,0,360)
\Line(70,50)(80,50)
\Line(30,50)(20,50)
\Line(50,30)(50,70)
\Line(30,50)(50,30)
\Line(70,50)(50,70)
\put(33,14){$M_{22},\,\ep^4$}
\end{picture}
\end{center}
}
\end{minipage}
\hspace{-5mm}
=-\frac{ 1 }{4\,\varepsilon^{3}}-\frac{3}{2\,\varepsilon^{2}}-\frac{33}{4 \,\ep }-\frac{175}{4}
+10 \sbz \zeta_{3}
+ \,\ep \left( -\frac{1825}{8}
+\frac{113 \sbz \zeta_{3}}{2}
+15 \zeta_4 
\right)\nonumber \displaybreak[1] \\ & 
+\,\varepsilon^{2}\left( -\frac{18867}{16}
+\frac{1241 \sbz \zeta_{3}}{4}
+\frac{339 \zeta_4 }{4}
+185 \sbz \zeta_{5}
\right)\nonumber \displaybreak[1] \\ & 
+\,\varepsilon^{3}\left( -\frac{194015}{32}
+\frac{13425 \sbz \zeta_{3}}{8}
+\frac{3723 \zeta_4 }{8}
+1028 \sbz \zeta_{5}
-204 \,\zeta_{3}^2 
+\frac{875 \zeta_6 }{2}
\right)\nonumber \displaybreak[1] \\ & 
+\,\varepsilon^{4}\left( -\frac{1987331}{64}
+\frac{143605 \sbz \zeta_{3}}{16}
+\frac{40275 \zeta_4 }{16}
+5588 \sbz \zeta_{5}
  \right. \nonumber \\   & \hspace{1.4cm} \left.  
 -1131 \,\zeta_{3}^2 
+\frac{9715 \zeta_6 }{4}
-612 \,\zeta_{3}\, \zeta_4 
+\frac{13157 \,\zeta_{7}}{4}
\right)+\,{\cal O}(\varepsilon^5)
\label{m22},
\\ 
\displaybreak[1]
{}&
\begin{minipage}{2.7cm}{
\begin{center}
\begin{picture}(50,80)(35,10)
\CArc(50,50)(20,0,360)
\CArc(50,60)(10,0,360)
\Line(20,50)(80,50)
\put(33,14){$M_{26},\,\ep^4$}
\end{picture}
\end{center}
}
\end{minipage}
\hspace{-5mm}
=-\frac{ 1 }{8\,\varepsilon^{3}}-\frac{13}{16\,\varepsilon^{2}}-\frac{141}{32 \,\ep }-\frac{1393}{64}
+2 \sbz \zeta_{3}
+ \,\ep \left( -\frac{12997}{128}
+13 \sbz \zeta_{3}
+3 \zeta_4 
\right)\nonumber \displaybreak[1] \\ & 
+\,\varepsilon^{2}\left( -\frac{116697}{256}
+\frac{123 \sbz \zeta_{3}}{2}
+\frac{39 \zeta_4 }{2}
+24 \sbz \zeta_{5}
\right)\nonumber \displaybreak[1] \\ & 
+\,\varepsilon^{3}\left( -\frac{1019645}{512}
+\frac{907 \sbz \zeta_{3}}{4}
+\frac{369 \zeta_4 }{4}
+156 \sbz \zeta_{5}
+\frac{49 \,\zeta_{3}^2 }{2}
+55 \zeta_6 
\right)\nonumber \displaybreak[1] \\ & 
+\,\varepsilon^{4}\left( -\frac{8732657}{1024}
+\frac{4375 \sbz \zeta_{3}}{8}
+\frac{2721 \zeta_4 }{8}
+693 \sbz \zeta_{5}
+\frac{637 \,\zeta_{3}^2 }{4}
+\frac{715 \zeta_6 }{2}
+\frac{147 \,\zeta_{3}\, \zeta_4 }{2}
+\frac{2475 \,\zeta_{7}}{4}
\right)+\,{\cal O}(\varepsilon^5)
\label{m26},
\\ 
\displaybreak[1]
{}&
\begin{minipage}{2.7cm}{
\begin{center}
\begin{picture}(50,80)(35,10)
\CArc(40,50)(20,0,360)
\CArc(60,50)(20,0,360)
\Line(50,33)(50,67)
\Line(20,50)(10,50)
\Line(80,50)(90,50)
\put(33,14){$M_{27},\,\ep^4$}
\end{picture}
\end{center}
}
\end{minipage}
\hspace{-5mm}
\hspace{7mm}=\frac{ 1 }{48\,\varepsilon^{3}}+\frac{7}{96\,\varepsilon^{2}}+\frac{11}{192 \,\ep }-\frac{605}{384}
+\frac{7 \sbz \zeta_{3}}{6}
\nonumber \displaybreak[1] \\ & 
+ \,\ep \left( -\frac{13525}{768}
+\frac{49 \sbz \zeta_{3}}{12}
+\frac{7 \zeta_4 }{4}
\right)+\,\varepsilon^{2}\left( -\frac{208037}{1536}
+\frac{161 \sbz \zeta_{3}}{6}
+\frac{49 \zeta_4 }{8}
+\frac{221 \sbz \zeta_{5}}{4}
\right)\nonumber \displaybreak[1] \\ & 
+\,\varepsilon^{3}\left( -\frac{2760397}{3072}
+\frac{9535 \sbz \zeta_{3}}{48}
+\frac{161 \zeta_4 }{4}
+\frac{1547 \sbz \zeta_{5}}{8}
-\frac{145 \,\zeta_{3}^2 }{3}
+\frac{3245 \zeta_6 }{24}
\right)\nonumber \displaybreak[1] \\ & 
+\,\varepsilon^{4}\left( -\frac{33789053}{6144}
+\frac{8273 \sbz \zeta_{3}}{6}
+\frac{9535 \zeta_4 }{32}
+\frac{14527 \sbz \zeta_{5}}{16}
  \right. \nonumber \\   & \hspace{1.4cm} \left.  
 -\frac{1015 \,\zeta_{3}^2 }{6}
+\frac{22715 \zeta_6 }{48}
-145 \,\zeta_{3}\, \zeta_4 
+\frac{11289 \,\zeta_{7}}{8}
\right)+\,{\cal O}(\varepsilon^5)
\label{m27},
\\ 
\displaybreak[1]
{}&
\begin{minipage}{2.7cm}{
\begin{center}
\begin{picture}(50,80)(35,10)
\CArc(20,50)(10,0,360)
\CArc(40,50)(10,0,360)
\CArc(66,50)(15,0,360)
\Line(10,50)(0,50)
\Line(50,50)(92,50)
\put(33,14){$M_{23},\,\ep^4$}
\end{picture}
\end{center}
}
\end{minipage}
\hspace{-5mm}
\hspace{7mm}=-\frac{ 1 }{4\,\varepsilon^{3}}-\frac{5}{8\,\varepsilon^{2}}-\frac{27}{16 \,\ep }-\frac{153}{32}
+\frac{3 \sbz \zeta_{3}}{2}
+ \,\ep \left( -\frac{891}{64}
+\frac{15 \sbz \zeta_{3}}{4}
+\frac{9 \zeta_4 }{4}
\right)\nonumber \displaybreak[1] \\ & 
+\,\varepsilon^{2}\left( -\frac{5265}{128}
+\frac{81 \sbz \zeta_{3}}{8}
+\frac{45 \zeta_4 }{8}
+\frac{21 \sbz \zeta_{5}}{2}
\right)\nonumber \displaybreak[1] \\ & 
+\,\varepsilon^{3}\left( -\frac{31347}{256}
+\frac{459 \sbz \zeta_{3}}{16}
+\frac{243 \zeta_4 }{16}
+\frac{105 \sbz \zeta_{5}}{4}
-\frac{9 \,\zeta_{3}^2 }{2}
+\frac{45 \zeta_6 }{2}
\right)\nonumber \displaybreak[1] \\ & 
+\,\varepsilon^{4}\left(
\boxeD{
 -\frac{187353}{512}
+\frac{2673 \sbz \zeta_{3}}{32}
+\frac{1377 \zeta_4 }{32}
+\frac{567 \sbz \zeta_{5}}{8}
-\frac{45 \,\zeta_{3}^2 }{4}
+\frac{225 \zeta_6 }{4}
-\frac{27 \,\zeta_{3}\, \zeta_4 }{2}
+\frac{147 \,\zeta_{7}}{2}
}
\right)
+\,{\cal O}(\varepsilon^5)
\label{m23},
\\ 
\displaybreak[1]
{}&
\begin{minipage}{2.7cm}{
\begin{center}
\begin{picture}(50,80)(35,10)
\CArc(40,50)(10,0,360)
\CArc(66,50)(15,0,360)
\Line(66,66)(82,50)
\Line(66,66)(50,50)
\Line(30,50)(20,50)
\Line(82,50)(92,50)
\put(33,14){$M_{24},\,\ep^4$}
\end{picture}
\end{center}
}
\end{minipage}
\hspace{-5mm}
\hspace{7mm}=-\frac{ 1 }{3\,\varepsilon^{3}}-\frac{4}{3\,\varepsilon^{2}}-\frac{16}{3 \,\ep }-\frac{64}{3}
+\frac{16 \sbz \zeta_{3}}{3}
+ \,\ep \left( -\frac{256}{3}
+\frac{64 \sbz \zeta_{3}}{3}
+8 \zeta_4 
\right)\nonumber \displaybreak[1] \\ & 
+\,\varepsilon^{2}\left( -\frac{1024}{3}
+\frac{256 \sbz \zeta_{3}}{3}
+32 \zeta_4 
+64 \sbz \zeta_{5}
\right)\nonumber \displaybreak[1] \\ & 
+\,\varepsilon^{3}\left( -\frac{4096}{3}
+\frac{1024 \sbz \zeta_{3}}{3}
+128 \zeta_4 
+256 \sbz \zeta_{5}
-\frac{128 \,\zeta_{3}^2 }{3}
+\frac{440 \zeta_6 }{3}
\right)\nonumber \displaybreak[1] \\ & 
+\,\varepsilon^{4}\left( -\frac{16384}{3}
+\frac{4096 \sbz \zeta_{3}}{3}
+512 \zeta_4 
+1024 \sbz \zeta_{5}
-\frac{512 \,\zeta_{3}^2 }{3}
+\frac{1760 \zeta_6 }{3}
-128 \,\zeta_{3}\, \zeta_4 
+768 \,\zeta_{7}
\right)+\,{\cal O}(\varepsilon^5)
\label{m24},
\\ 
\displaybreak[1]
{}&
\begin{minipage}{2.7cm}{
\begin{center}
\begin{picture}(50,80)(35,10)
\CArc(50,50)(20,0,360)
\Line(31,45)(21,45)
\Line(69,45)(79,45)
\Line(38,66)(62,66)
\Line(38,66)(31,45)
\Line(62,66)(69,45)
\put(33,14){$M_{25},\,\ep^4$}
\end{picture}
\end{center}
}
\end{minipage}
\hspace{-5mm}
=-\frac{3}{8\,\varepsilon^{3}}-\frac{33}{16\,\varepsilon^{2}}-\frac{345}{32 \,\ep }-\frac{3525}{64}
+\frac{45 \sbz \zeta_{3}}{4}
+ \,\ep \left( -\frac{35625}{128}
+\frac{495 \sbz \zeta_{3}}{8}
+\frac{135 \zeta_4 }{8}
\right)\nonumber \displaybreak[1] \\ & 
+\,\varepsilon^{2}\left( -\frac{358125}{256}
+\frac{5175 \sbz \zeta_{3}}{16}
+\frac{1485 \zeta_4 }{16}
+\frac{855 \sbz \zeta_{5}}{4}
\right)\nonumber \displaybreak[1] \\ & 
+\,\varepsilon^{3}\left( -\frac{3590625}{512}
+\frac{52875 \sbz \zeta_{3}}{32}
+\frac{15525 \zeta_4 }{32}
+\frac{9405 \sbz \zeta_{5}}{8}
-\frac{675 \,\zeta_{3}^2 }{4}
+\frac{2025 \zeta_6 }{4}
\right)\nonumber \displaybreak[1] \\ & 
+\,\varepsilon^{4}\left( -\frac{35953125}{1024}
+\frac{534375 \sbz \zeta_{3}}{64}
+\frac{158625 \zeta_4 }{64}
+\frac{98325 \sbz \zeta_{5}}{16}
  \right. \nonumber \\   & \hspace{1.4cm} \left.  
 -\frac{7425 \,\zeta_{3}^2 }{8}
+\frac{22275 \zeta_6 }{8}
-\frac{2025 \,\zeta_{3}\, \zeta_4 }{4}
+\frac{16245 \,\zeta_{7}}{4}
\right)+\,{\cal O}(\varepsilon^5)
\label{m25},
\\ 
\displaybreak[1]
{}&
\begin{minipage}{2.7cm}{
\begin{center}
\begin{picture}(50,80)(35,10)
\CArc(40,50)(10,0,360)
\CArc(60,53)(10,0,360)
\CArc(60,47)(10,0,360)
\Line(30,50)(20,50)
\Line(70,50)(80,50)
\put(33,14){$M_{11},\,\ep^5$}
\end{picture}
\end{center}
}
\end{minipage}
\hspace{-5mm}
=\frac{ 1 }{36\,\varepsilon^{2}}+\frac{35}{216 \,\ep }+\frac{991}{1296}+ \,\ep \left( \frac{26207}{7776}
-\frac{11 \sbz \zeta_{3}}{18}
\right)+\,\varepsilon^{2}\left( \frac{670951}{46656}
-\frac{385 \sbz \zeta_{3}}{108}
-\frac{11 \zeta_4 }{12}
\right)\nonumber \displaybreak[1] \\ & 
+\,\varepsilon^{3}\left( 
\boxeD{
\frac{16852031}{279936}
-\frac{10901 \sbz \zeta_{3}}{648}
-\frac{385 \zeta_4 }{72}
-\frac{13 \sbz \zeta_{5}}{2}
}
\right)\nonumber \displaybreak[1] \\ & 
+\,\varepsilon^{4}\left(
\boxeD{
 \frac{417941623}{1679616}
-\frac{288277 \sbz \zeta_{3}}{3888}
-\frac{10901 \zeta_4 }{432}
-\frac{455 \sbz \zeta_{5}}{12}
+\frac{121 \,\zeta_{3}^2 }{18}
-\frac{265 \zeta_6 }{18}
}
\right)\nonumber \displaybreak[1] \\ & 
+\,\varepsilon^{5}\left(
\boxeD{
 \frac{10274059439}{10077696}
-\frac{7380461 \sbz \zeta_{3}}{23328}
-\frac{288277 \zeta_4 }{2592}
-\frac{12883 \sbz \zeta_{5}}{72}
}
  \right. \nonumber \\   & \hspace{1.4cm} \left.  
\boxeD{
 +\frac{4235 \,\zeta_{3}^2 }{108}
-\frac{9275 \zeta_6 }{108}
+\frac{121 \,\zeta_{3}\, \zeta_4 }{6}
-\frac{433 \,\zeta_{7}}{6}
}
\right)+\,{\cal O}(\varepsilon^6)
\label{m11},
\\ 
\displaybreak[1]
{}&
\begin{minipage}{2.7cm}{
\begin{center}
\begin{picture}(50,80)(35,10)
\CArc(34,50)(15,0,360)
\CArc(66,50)(15,0,360)
\Line(10,50)(90,50)
\put(33,14){$M_{12},\,\ep^5$}
\end{picture}
\end{center}
}
\end{minipage}
\hspace{-5mm}
\hspace{7mm}=\frac{ 1 }{16\,\varepsilon^{2}}+\frac{5}{16 \,\ep }+\frac{79}{64}+ \,\ep \left( \frac{9}{2}
-\frac{3 \sbz \zeta_{3}}{4}
\right)+\,\varepsilon^{2}\left( \frac{4041}{256}
-\frac{15 \sbz \zeta_{3}}{4}
-\frac{9 \zeta_4 }{8}
\right)\nonumber \displaybreak[1] \\ & 
+\,\varepsilon^{3}\left( \frac{13851}{256}
-\frac{237 \sbz \zeta_{3}}{16}
-\frac{45 \zeta_4 }{8}
-\frac{21 \sbz \zeta_{5}}{4}
\right)+\,\varepsilon^{4}\left(
\boxeD{
 \frac{186867}{1024}
-54 \sbz \zeta_{3}
-\frac{711 \zeta_4 }{32}
-\frac{105 \sbz \zeta_{5}}{4}
+\frac{9 \,\zeta_{3}^2 }{2}
-\frac{45 \zeta_6 }{4}
}
\right)\nonumber \displaybreak[1] \\ & 
+\,\varepsilon^{5}\left( 
\boxeD{
\frac{311283}{512}
-\frac{12123 \sbz \zeta_{3}}{64}
-81 \zeta_4 
-\frac{1659 \sbz \zeta_{5}}{16}
+\frac{45 \,\zeta_{3}^2 }{2}
-\frac{225 \zeta_6 }{4}
+\frac{27 \,\zeta_{3}\, \zeta_4 }{2}
-\frac{147 \,\zeta_{7}}{4}
}
\right)+\,{\cal O}(\varepsilon^6)
\label{m12},
\\ 
\displaybreak[1]
{}&
\begin{minipage}{2.7cm}{
\begin{center}
\begin{picture}(50,80)(35,10)
\CArc(50,50)(20,0,360)
\CArc(65,58)(11,0,360)
\Line(58,68)(30,49.3333333333333)
\Line(30,49)(20,49)
\Line(70,47)(80,47)
\put(33,14){$M_{13},\,\ep^5$}
\end{picture}
\end{center}
}
\end{minipage}
\hspace{-5mm}
=\frac{ 1 }{32\,\varepsilon^{2}}+\frac{5}{24 \,\ep }+\frac{1309}{1152}+ \,\ep \left( \frac{317}{54}
-\frac{9 \sbz \zeta_{3}}{8}
\right)+\,\varepsilon^{2}\left( \frac{1234309}{41472}
-\frac{15 \sbz \zeta_{3}}{2}
-\frac{27 \zeta_4 }{16}
\right)\nonumber \displaybreak[1] \\ & 
+\,\varepsilon^{3}\left( \frac{4658207}{31104}
-\frac{1309 \sbz \zeta_{3}}{32}
-\frac{45 \zeta_4 }{4}
-\frac{153 \sbz \zeta_{5}}{8}
\right)\nonumber \displaybreak[1] \\ & 
+\,\varepsilon^{4}\left( \frac{1121384029}{1492992}
-\frac{634 \sbz \zeta_{3}}{3}
-\frac{3927 \zeta_4 }{64}
-\frac{255 \sbz \zeta_{5}}{2}
+\frac{81 \,\zeta_{3}^2 }{4}
-45 \zeta_6 
\right)\nonumber \displaybreak[1] \\ & 
+\,\varepsilon^{5}\left( \frac{2105747071}{559872}
-\frac{1234309 \sbz \zeta_{3}}{1152}
-317 \zeta_4 
-\frac{22253 \sbz \zeta_{5}}{32}
  \right. \nonumber \\   & \hspace{1.4cm} \left.  
 +135 \,\zeta_{3}^2 
-300 \zeta_6 
+\frac{243 \,\zeta_{3}\, \zeta_4 }{4}
-\frac{2781 \,\zeta_{7}}{8}
\right)+\,{\cal O}(\varepsilon^6)
\label{m13},
\\ 
\displaybreak[1]
{}&
\begin{minipage}{2.7cm}{
\begin{center}
\begin{picture}(50,80)(35,10)
\CArc(50,50)(20,0,360)
\CArc(40,50)(10,0,360)
\CArc(60,50)(10,0,360)
\Line(30,50)(20,50)
\Line(70,50)(80,50)
\put(33,14){$M_{14},\,\ep^5$}
\end{picture}
\end{center}
}
\end{minipage}
\hspace{-5mm}
=\frac{ 1 }{24\,\varepsilon^{2}}+\frac{49}{144 \,\ep }+\frac{1867}{864}+ \,\ep \left( \frac{64813}{5184}
-\frac{23 \sbz \zeta_{3}}{12}
\right)\nonumber \displaybreak[1] \\ & 
+\,\varepsilon^{2}\left( \frac{2146387}{31104}
-\frac{1127 \sbz \zeta_{3}}{72}
-\frac{23 \zeta_4 }{8}
\right)+\,\varepsilon^{3}\left( \frac{69116413}{186624}
-\frac{42941 \sbz \zeta_{3}}{432}
-\frac{1127 \zeta_4 }{48}
-\frac{127 \sbz \zeta_{5}}{4}
\right)\nonumber \displaybreak[1] \\ & 
+\,\varepsilon^{4}\left( \frac{2185200787}{1119744}
-\frac{1490699 \sbz \zeta_{3}}{2592}
-\frac{42941 \zeta_4 }{288}
-\frac{6223 \sbz \zeta_{5}}{24}
+\frac{529 \,\zeta_{3}^2 }{12}
-\frac{895 \zeta_6 }{12}
\right)\nonumber \displaybreak[1] \\ & 
+\,\varepsilon^{5}\left( \frac{68213322013}{6718464}
-\frac{49366901 \sbz \zeta_{3}}{15552}
-\frac{1490699 \zeta_4 }{1728}
-\frac{237109 \sbz \zeta_{5}}{144}
  \right. \nonumber \\   & \hspace{1.4cm} \left.  
 +\frac{25921 \,\zeta_{3}^2 }{72}
-\frac{43855 \zeta_6 }{72}
+\frac{529 \,\zeta_{3}\, \zeta_4 }{4}
-\frac{2189 \,\zeta_{7}}{4}
\right)+\,{\cal O}(\varepsilon^6)
\label{m14},
\\ 
\displaybreak[1]
{}&
\begin{minipage}{2.7cm}{
\begin{center}
\begin{picture}(50,80)(35,10)
\CArc(50,54)(15,0,360)
\CArc(50,46)(15,0,360)
\Line(20,50)(80,50)
\put(33,14){$M_{01},\,\ep^6$}
\end{picture}
\end{center}
}
\end{minipage}
\hspace{-5mm}
=-\frac{ 1 }{576 \,\ep }-\frac{13}{768}-\frac{9823}{82944}\,\varepsilon+\,
\varepsilon^{2}\left( 
\boxeD{
-\frac{80513}{110592}
+\frac{13 \sbz \zeta_{3}}{144}
}
\right)\nonumber \displaybreak[1] \\ & 
+\,\varepsilon^{3}\left(
\boxeD{
 -\frac{49995799}{11943936}
+\frac{169 \sbz \zeta_{3}}{192}
+\frac{13 \zeta_4 }{96}
}
\right)+\,\varepsilon^{4}\left(
\boxeD{
 -\frac{122739515}{5308416}
+\frac{127699 \sbz \zeta_{3}}{20736}
+\frac{169 \zeta_4 }{128}
+\frac{67 \sbz \zeta_{5}}{48}
}
\right)\nonumber \displaybreak[1] \\ & 
+\,\varepsilon^{5}\left(
\boxeD{
 -\frac{213973312663}{1719926784}
+\frac{1046669 \sbz \zeta_{3}}{27648}
+\frac{127699 \zeta_4 }{13824}
+\frac{871 \sbz \zeta_{5}}{64}
-\frac{169 \,\zeta_{3}^2 }{72}
+\frac{235 \zeta_6 }{72}
}
\right)\nonumber \displaybreak[1] \\ & 
+\,\varepsilon^{6}\left(
\boxeD{
 -\frac{1507417628113}{2293235712}
+\frac{649945387 \sbz \zeta_{3}}{2985984}
+\frac{1046669 \zeta_4 }{18432}
+\frac{658141 \sbz \zeta_{5}}{6912}
}
 \right. \nonumber \\   & \hspace{1.4cm} \left.
\boxeD{
 -\frac{2197 \,\zeta_{3}^2 }{96}
+\frac{3055 \zeta_6 }{96}
-\frac{169 \,\zeta_{3}\, \zeta_4 }{24}
+\frac{373 \,\zeta_{7}}{16}
}
\right)+\,{\cal O}(\varepsilon^7)
\label{m10},
\\ 
\displaybreak[1]
{}&
\begin{minipage}{2.7cm}{
\begin{center}
\begin{picture}(50,80)(35,10)
\CArc(20,50)(10,0,360)
\CArc(40,50)(10,0,360)
\CArc(60,50)(10,0,360)
\CArc(80,50)(10,0,360)
\Line(10,50)(0,50)
\Line(90,50)(100,50)
\put(33,14){$M_{31},\,\ep^3$}
\end{picture}
\end{center}
}
\end{minipage}
\hspace{-5mm}
\hspace{7mm}=\boxeD{\frac{1}{\,\varepsilon^{4}}}+\,{\cal O}(\varepsilon^4)
\label{m31}
{}.
\end{align}

\ice{

M[23, 4]

M[12, 4]
M[12, 5]

M[11, 3]
M[11, 4]
M[11, 5]

M[10, 2]
M[10, 3]
M[10, 4]
M[10, 5]
M[10, 6]
}

Thus, we observe that at the four-loop level the GaC method works as
good as the three-loop ones: {\bf all } required terms of the
$\ep$-expansion of every four-loop MI have been expressed in terms of
only {\bf twelve}  coefficients (boxed in   eqs.~(\ref{m61}-\ref{m31}))
\beq
M_{23,4}, M_{11,3}, M_{11,4}, M_{11,5},  M_{12,4}, M_{12,5},
M_{01,2}, M_{01,3},M_{01,4},M_{01,5},M_{01,6},M_{31,-4},
\label{all_12}
\eeq
of {\bf primitive  watermelon-like}  massless propagator integrals.

An inspection of the above  results for MI's   reveals a few remarkable features.
\begin{enumerate}
%
\item 
In agreement with common expectations (based on the known solutions of the two- and three-loop B-problem) the 
transcendental terms up to (and including) weight 7 appear  in eqs.~(\ref{m61}-\ref{m10}).  
That is all results depend  on only {\em } five irrational constants: $\zeta_3$, $\zeta_4$, $\zeta_5$, $\zeta_6$ and $\zeta_7$.

\item For a given MI $M_i$ the  term $\ep^{p_i}$ (that is one  with  maximal power in $\ep$) {\em always}\footnote{
MI $M_{31}$ is the only exception from this rule since its sub-leading in $\ep$ terms are fixed to be zero essentially 
by hands, that is  by choosing the G-scheme.}
      includes  $\zeta_7$.   

\item A  term proportional to  $\ep^{p_i -j}$ could contain $\zeta_n$ with $n$  not exceeding $7-j$;
if $ 7 - j < 3$ then the term is free from irrational numbers.

\item There is another restriction on  the singular part of any MI's
(in fact, it is valid for an arbitrary p-intergral).  It states that the
term $\ep^{-n}$ (with $n=1,2,3,4$) may not contain zetas with the
transcendentally weight  exceeding $(7-2n)$.  This property explains a very peculiar feature of  MI's 
$M_{62}$ and $M_{63}$:
the absence of  $\zeta_6$ in the  corresponding  ${\cal O}(\ep^{p_i -1})$ terms.

\item The only two {\em finite} MI's, namely, $M_{44}$ and $M_{45}$ contain only  terms of one and same 
weight in every (available) coefficient of their $\ep$-expansions. 

\item The same property of "transcendental homogeneity" is true for
the MI $M_{52}$ (which is up to a factor of $1/\ep$ is the three-loop {\em
finite} MI $N_0$) {\em if} one divides an extra factor $(1-2\ep)^2$ out
of it.  (See in this connection work \cite{Broadhurst:1999xk}, where some general 
arguments were given in favour of the hypothesis that the property is
valid in all orders in $\ep$.)

\end{enumerate}

We want to stress that any statement on the structure of $\zeta$'s  appearing in an integral does  depend
on the  global normalization which is rather arbitrary. Our normalization condition is a  natural one 
but, certainly, not unique. If we would choose
\beq
M_{31} = \frac{1}{\ep^4}(1 + \sum_{1 \le i \le 7} a_i\ep^i) 
\label{M31}
\eeq
then  all MI's would   depend on $a_i$, and all statements just discussed   above  could be, obviously, made  invalid
{\em if} a coefficient $a_i$ were   allowed to contain $\zeta_i$ (for $i > 1$) and $\gamma_E$ for $i=1$.

On the other hand, if we restrict ourselves to  a natural choice of the   normalization of MI $M_{31}$ such as
\beq
M_{31} = \frac{1}{\ep^4}(1 +  \sum_{3 \le i \le 7} b_i \zeta_i \ep^i) 
\label{M31_nat}
{},
\eeq
with $b_i$ being rational numbers, then the properties 1-6 would in  general stay untouched.

\subsection{Tests of the results \label{test:all}}

In this subsection we discuss various checks which  we have  made to
test our results expressed in eqs.~(\ref{m61}-\ref{m31}).
The set of 28 master integrals is  naturally divided in three  subsets:
primitive\\ 
$( 
M_{23}\,,M_{24}\,,M_{25}\,,M_{11}\,,M_{12}\,,M_{13}\,,M_{14}\,,M_{01}\,,M_{31})$,
simple $(
M_{32}\,,M_{33}\,,M_{21}\,,M_{22}\,,M_{26}\,,M_{27}
)$
and, finally,  complicated ones
$( M_{61}\,,M_{62}\,,M_{63}\,,M_{51}\,,M_{41}\,,M_{42}\,,M_{44}\,,M_{45}\,, M_{34}\,, M_{35}\,,M_{36}\,,M_{52}
\,,M_{43})$. We will consider these  subsets  separately. 

\subsubsection{Primitive integrals \label{test:primitive}}
A primitive FI is by definition expressible in terms of the $\Gamma$-function. A straightforward use of formulas
of section~\ref{G-functions} gives:

\bea
M_{23} &=& \frac{1}{\ep^3}\,G(\ep,1), \ \ \ \
M_{24} = \frac{1}{\ep^3}\,G(2\ep,1), \ \ \ \
M_{25} = \frac{1}{\ep^3}\,G(3\ep,1), \ \ \ \
\nn
\\
M_{11} &=& \frac{1}{\ep^3}\,G(\ep,\ep), \ \ \ \
M_{12} = \frac{1}{\ep^2}\,G(\ep,1)\,G(\ep,1), \ \ \ \
M_{13} = \frac{1}{\ep^2}\,G(\ep,1)\,G(-1+3\ep,1),
\nn
\\
M_{14} &=& \frac{1}{\ep^3}\,G(\ep,2\ep), \ \ \ \
M_{01} = \frac{1}{\ep^2}\,G(\ep,\ep)\,G(-2+3\ep,1) , \ \ \ \
M_{31} = \frac{1}{\ep^4} 
{}.
\label{simple}
\eea

After the expansion in $\ep$ eqs.~\re{simple} produce altogether 
$89$ coefficients.
As was discussed in subsection \ref{sol4} as many as  twelve 
coefficients listed in eq.~\re{all_12}
have been used in the  process
of the solution of the system of GaC equations, while the remaining 77  coefficients
have been {\em predicted} from the equation and listed unboxed in  
eqs.~(\ref{m23}--\ref{m10}). 

The  reader is advised to check that all these 77 coefficients are in full agreement to 
eqs. (\ref{m23}--\ref{m10}).

\subsubsection{Simple  integrals \label{test:simple}}

These all could  be expressed in terms of $G$-functions and the generalized two-loop diagram:
\bea
M_{32} &=& \frac{1}{\ep^2}\,F(1,1,1,1,\ep)
\label{sm32}
{},
\\
M_{33} &=& \frac{1}{\ep^2}\,F(1,1,1,1,2\ep)
\label{sm33}
\\
M_{21} &=& \frac{1}{\ep^2}\,F(1,1,1,\ep,\ep)
\label{sm21}
{},
\\
M_{22} &=& \frac{1}{\ep^2}\,F(1,\ep,1,\ep,1)
\label{sm22}
{},
\\
M_{26} &=& \frac{1}{\ep}\,G(3\ep,1)\,F(1,1,1,1,\ep)
\label{sm26}
{},
\\
M_{27} &=& \frac{1}{\ep}\,G(\ep,1)\,F(1,1,1,1,2\ep -1)
\label{sm27}
{}.
\eea
%
To be specific, let us consider the direct evaluation of  $M_{33}$ in some details.
First, we define a related FI  $M'_{33}$  pictured in   eq.~ \re{m33p} below.
A simple reduction of FI  $M'_{33}$   to MI's 
gives the (exact) equation 
\beq
\begin{minipage}{2.7cm}{
\begin{center}
\begin{picture}(50,80)(35,0)
\SetScale{0.8}
\SetWidth{1.0}
\CArc(50,50)(20,0,360)
\CArc(50,64)(6,0,360)
\CArc(50,52)(6,0,360)
\Line(50,30)(50,45)
\Line(30,50)(20,50)
\Line(70,50)(80,50)
\put(30,10){$M'_{33}$}
\end{picture}
\end{center}
}
\end{minipage}
\hspace{-10mm}
 = \hspace{5mm} \, \frac{4 (7 D - 26) (5 D - 14) (5 D - 16) (D - 3)}{9 (3 D - 10) (D - 4)^3}\, M_{14}
+
\frac{-2 (2 D - 7)}{3 (D - 4)} 
\,M_{33}
{}.
\label{m33p}
\eeq
On the other hand, 
\beq
M'_{33} = \frac{1}{\ep^2} \,F(1,1,1,1,1+2\ep)
{}
\label{m33pF}
\eeq
and  (see \cite{Kazakov:1983pk} as well as \cite{Broadhurst:1986bx,Barfoot:1987kg,Bierenbaum:2003ud}) 
\bea
&{}& \hspace{-2cm}\frac{1}{\left( 1-2\ep \right)}
F(1,1,1,1,1+2\ep)
=
 6 \zeta_3+9 \zeta_4 \ep+192 \zeta_5 \ep^2
 +\left(465 \zeta_6-168 \zeta_3^2 \right) \ep^3
 \nonumber 
\\
 &{} &
 +\left(4509 \zeta_7-504 \zeta_4 \zeta_3 \right) \ep^4
 +\left(\frac{16377}{2} \zeta_8 -1620 \zeta_{6,2}-3252 \zeta_5 \zeta_3  \right) \ep^5
 \nonumber 
\\
 & &
 +\left(98490 \zeta_9-14598 \zeta_5 \zeta_4-15390 \zeta_6 \zeta_3  +2676\zeta_3^3 \right) \ep^6
 + {\cal O}(\ep^7)
{},
\label{F}
\eea
where 
\[\zeta_{6,2} \equiv \sum_{n_1 > n_2 > 0} \frac{1}{n_1^6\, n_2^2} .\]

Finally, eqs.~(\ref{m33p}-\ref{F}) together with eq.~\re{m14} lead to the following ({\em independent} from our calculations) result
for MI $M_{33}$ which is not only in full agreement to eq.~\re{m33} but also includes 
two more terms in $\ep$:
\bea 
M_{33,4}  &=&   -\frac{367253}{3} + \frac{97982}{3} \zeta_3 - \frac{11038}{3}\zeta_3^2 + 7987 \zeta_4 -
     1966 \zeta_3 \zeta_4 
\nn
\\
&+&  22750 \zeta_5 - 3914 \zeta_3 \zeta_5 + \frac{31690}{3} \zeta_6
    - 4860 \zeta_{6,2} + 22676 \zeta_7 + \frac{147181}{8} \zeta_8
\label{m33,4}
{},
\\ 
M_{33,5}  &=& 
 -\frac{2073833}{3} + \frac{580022}{3} \zeta_3 - \frac{66370}{3} \zeta_3^2 + \frac{47918}{9} \zeta_3^3 +
     48991 \zeta_4 - 11038 \zeta_3 \zeta_4 
\nn
\\   
&+&  123766 \zeta_5 - 7828 \zeta_3 \zeta_5 -
     35031 \zeta_4 \zeta_5 + \frac{164350}{3} \zeta_6  
\nn
\\ 
&-&
    \frac{97340}{3} \zeta_3 \zeta_6 - 9720 \zeta_{6,2} +
     103838 \zeta_7 + \frac{147181}{4} \zeta_8 + \frac{2293555}{9} \zeta_9
\label{m33,5}
{}.
\eea
\noindent

\ice{
m33extep4 = -367253/3 + (97982*z3)/3 - (11038*z3^2)/3 + 7987*z4 -
     1966*z3*z4 + 22750*z5 - 3914*z3*z5 + (31690*z6)/3 - 4860*z62 +
     22676*z7 + (147181*z8)/8

m33extep5 = -2073833/3 + (580022*z3)/3 - (66370*z3^2)/3 + (47918*z3^3)/9 +
     48991*z4 - 11038*z3*z4 + 123766*z5 - 7828*z3*z5 - 35031*z4*z5 +
     (164350*z6)/3 - (97340*z3*z6)/3 - 9720*z62 + 103838*z7 + (147181*z8)/4 +
     (2293555*z9)/9

}

In the same way we have successfully checked all other simple
MI's. In addition, for  all of them we  get two extra terms of the $\ep$-expansion. 
They look similar to the ones listed in (\ref{m33,4},\ref{m33,5}) and include, in addition to 
$\zeta_3 - \zeta_9$, only $\zeta_{6,2}$.

\subsubsection{Complicated   integrals \label{test:complicated} }

We start from diagrams $M_{52}$ and $M_{43}$ which are relatively
simple as they could be expressed through the $\ep^1$ and $\ep^2$ extra
terms of the basic three-loop non-planar integral $N_0$, namely:
\bea
M_{52} &=& \frac{1}{\ep} N_0(\ep)  = 
\frac{20\zeta_5}{\ep} + N_{0,1} +  N_{0,2}\,\ep + {\cal O}(\ep^2),
\\
M_{43} &=& G(1,2+3\ep) \,N_0(\ep) =- \frac{5 \zeta_5}{\ep} +  + 25 \zeta_5 - \frac{N_{0,1}}{4}
+( \frac{6 N_{0,1}}{4} -  \frac{N_{0,2}}{4} -75\zeta_5) \ep 
  +{\cal O}(\ep^2)
{}.
\eea
Thus,
\beq
N_{0,1} =   -80 \sbz \zeta_{5} +68 \,\zeta_{3}^2  +50 \zeta_6 
, \ \ 
N_{0,2} =-272 \,\zeta_{3}^2 
-200 \zeta_6 
+204 \,\zeta_{3}\, \zeta_4 
+450 \,\zeta_{7}
\label{N0:1and2}
{}.
\eeq
\ice{
[12]:=  Coll[ Exn[Ex[Exn[G[1,2+3*ep] + O[ep]^3]*(20*z5 +N1*ep+N2*ep^2)] +O[ep]^2 ],ep]//InputForm

Out[12]//InputForm= -N1/4 + ep*((5*N1)/4 - N2/4 - 75*z5) + 25*z5 - (5*z5)/ep
}
The coefficient  $N_{0,1}$ was known since long from calculations of
the five-loop $\beta$-function in the $\phi^4$-model in
\cite{Kazakov:uniqueness:PRB:83}.  The second coefficient $N_{0,2}$
was first computed with the GaC method and presented in
\cite{Chetyrkin:Talk:LL04}. Its completely independent calculation
(through fitting a high-precision numerical result with an appropriate
analytical ansatz) was performed in \cite{Bekavac:2005xs}. Needless to
say that the results of \cite{Kazakov:uniqueness:PRB:83} and
\cite{Bekavac:2005xs} are in agreement with eq.~\re{N0:1and2}.

All other complicated integrals (except for convergent integral
$M_{44}$, whose value,  $\frac{447}{8} \zeta_7$, at $D=4$ was
also analytically found in \cite{Kazakov:uniqueness:PRB:83}) have not
been known with sufficient accuracy  before
our calculations.  Note, that for a given (four-loop) master integral
$M_{i}$ one needs to know only the  $5+p_i$ first terms in its
$\ep$-expansion.  (For accounting purposes we assume that every
expansion starts from $\frac{1}{\ep^4}$ even if the corresponding term
drops out from a specific MI.)  
Among them first $5+p^i-1$ (that is
all except for the last one) are in a sense easy as they all could be
analytically found by well-known methods based on Infrared
Rearrangement (see, e.g.  \cite{Chetyrkin:2006dh} where the issue
was spelled out on the example of massive four-loop tadpoles).
At any case they all are very well checked  in the course of  the renormalization procedure.


\section{Perspectives \label{perspectives} }

\subsection{ Five-loop master integrals \label{perspectives:Bprb}}

As we have seen from the discussion in the previous section  all complicated and even simple four-loop MI's 
have been completely  expressed in terms of very simple watermelon-like primitive p-integrals. In fact, the reduction 
method based on the $1/D$ expansion of coefficient functions of MI's is in  general applicable for {\bf }
any number of loops in the  A-Problem. The  GaC symmetry is also  not limited by number of loops of p-integrals.

Thus, if the five-loop A-problem is  solved,  then the five-loop B-problem can also be solved in the  following  sense:
the identities stemming  from the GaC symmetry will express all five-loop MI's in terms of significantly smaller 
set  of p-integrals.  But which exactly set?  At present nobody knows for sure. But one could certainly expect that:

\begin{itemize}
\item
in general the  five-loop master p-integrals   will contain irrational terms of weight not higher than 9;

\item  
the "small set" of five-loop integrals will include ones  primitive as well as those expressible  in terms
of the generalized  F-function.  

\end{itemize}

As both types of the  integrals could  certainly be  analytically  evaluated up to 
the weight 9 \cite{Chetyrkin:1980pr,Barfoot:1987kg}
we conclude that    the five-loop B-problem should be analytically doable. Moreover, we believe that the GaC symmetry + reduction
provides the simplest way of analytical solution of the five-loop B-problem.


\subsection{General  case \label{perspectives:general}}

At first sight the applicability scope of the GaC method is rather
limited and  amounts exclusively to the massless propagators. Indeed, the
heart of the method is the existence of relations between integrals of
{\em different} topologies {\em beyond} those provided by the very
integration by parts. We are not aware about existence of such
relations in general case except for a one: {\em finiteness}.

Indeed, two finite  integrals are in certainly equal to each
other with accuracy ${\cal O}(\ep^0)$ irrespectively on their
topologies. As a result IBP relations will provide some partial
information about the values of corresponding master integrals.
Unfortunately, the information proves to be rather  limited\footnote{
Nevertheless, there are cases for which even this 
limited  information is  enough, see e.g.  \cite{Czakon:2004bu,Eden:2009hz}.
}. 

Indeed, let us consider, as a simple example,  eqs.~\re{eqs_from_Gl:3lB}. Without any use of 
GaC symmetry one could, obviously,  write
\beq 
N_0  = L_0 +  {\cal O}(\ep^0), \ \ N_0  = N_1 +  {\cal O}(\ep^0), \ \  \ N_0  = N_2 +  {\cal O}(\ep^0),
 \  \ N_0  =  {\cal O}(\ep^0)
\label{eqs_from_finitness:3lB1}
{}.
\eeq
After reduction to masters and solution of the resulting equations we
arrive to the same results ~(\ref{N0res}-\ref{P2}) but with the  $\ep$-accuracy
{\em downgraded} by one for every master integral. This is certainly
{\em not enough} to solve the three-loop problem:  no new information is 
obtained for the most complicated non-planar master integral $N_0$.

The reason for the failure is quite clear: the equations  \re{eqs_from_Gl:3lB} do not provide, in fact,
any constraints  on the value of $N_0$ as they could be equivalently rewritten as follows:
\beq 
 L_0 =  {\cal O}(\ep^0), \ \  N_1 =  {\cal O}(\ep^0), \ \  \   N_2 =  {\cal O}(\ep^0), \  \ N_0  =  {\cal O}(\ep^0)
\label{eqs_from_finitness:3lB:2}
{}.
\eeq

Repeating the same exercise at four-loop level we will arrive to a
similar conclusion: without any use of the GaC symmetry one could find
for every master integral $M_i$ all except for the last one (that is
$5+p^i-1$) of its $\ep$-expansion. Again finiteness only, without the
GaC-symmetry, is not enough to solve the four-loop Problem. 

On the other hand, any L-loop MI multiplied by a one-loop scalar massless propagator 
is, obviously,  a (L+1)-loop MI (compare, for, example, $T_1$ and $P_2$).
Within the G-scheme  framework the values of the  integrals are trivially related
by  a factor $\ep$.
Thus, at two- and three-loops we get the following identities
\[
T_1 = \ep P_2,
\ \ 
T_2 = \ep P_4
\]
and
\[
N_0 = \ep M_{52},
\ \ 
L_1 = \ep M_{32},
\ \ 
P_1 = \ep M_{24},
\ \ 
P_2 = \ep M_{23},
\ \ 
P_3 = \ep M_{11}
{}
\]
respectively.

Using "downgraded" eqs.~\re{m52} and \re{m32} we find the same results
for both non-primitive MI's $N_0$ and $L_1$ as in eqs.~\re{N0res} and
\re{L1} but  with the deepness of the $\ep$-expansion {\em increased} by one.  Thus we arrive at
a truly remarkable conclusion: by merely reducing finite four-loop
propagators to the master integrals and {\em without} any use of GaC
symmetry we could not only completely solve the three-loop Problem, we  
even can  get one more term in $\ep$-expansion of every non-trivial  master
integral!

It remains to see how predictive is this trick of finding master
integrals for cases with other patterns of external momenta and
masses. But its is absolutely clear that at least some  useful nontrivial
information can be obtained along these lines.

\section{Even zetas  \label{evenz} } 

\subsection{Four and five loops \label{evenz:4and5}}

In addition to six  remarkable  features of  four-loop master p-integrals listed in subsection \ref{sol4}  
there exist  the seventh one,  probably most  remarkable. Indeed, a scrupulous  inspection of  eqs.~(\ref{m61}-\ref{m10}))
demonstrate that  all their right hand sides do depend on only the following {\bf three} combinations of zetas:
\beq
\hat{\zeta}_3 = \zeta_3+\frac{3 \ep}{2}{\zeta_4}-\frac{5 \ep^3}{2}{\zeta_6},
\,\, 
\hat{\zeta}_5=\zeta_5+\frac{5\ep}{2}\zeta_6 \ \ \  \mbox{and} \ \  \ \zeta_7.
\label{hat_zetas}
\eeq
This simple   fact has far reaching consequences. Indeed, a little meditation on \re{hat_zetas} leads to  the  following 
statement\footnote{By any p-integral below we understand any one with number of loops less or equal to four.}:

\noindent {\bf Theorem 4.}
\begin{enumerate}
\item Any  {\em finite} at $\ep \to 0 $  p-integral  does not contain even zetas $\{\zeta_{2n} \mid n \ge 2\}$ in the limit 
of $\ep \rightarrow 0$.

\item Any {\em finite} at $\ep \to 0 $ combination of p-integrals like
\[
 \sum C_i(\ep) p_i {}, \ \ C_i = \sum_j  C_{ij} \ep^j
{},
\]
with the coefficient functions being  functions (not necessarily  finite at  $\ep \to  0$)   
 with purely rational coefficients $C_{ij}$,  will  {\em not} contain
even zetas in the limit of $\ep \rightarrow 0$ (while  {\em odd}  zetas  \\
\mbox{  $\{\zeta_{2n+1} \mid n \ge 1\}$ } are expected and indeed  appear in general).

\item Let $F(\ep)$ be any renormalized (and, thus, finite in the limit
of $\ep \to 0$) combination of any p-integrals. The {\em sole} source of
possible even zetas  in $F(0)$ is   the appearance of  zetas (not necessarily even) in the
renormalization factors involved in carrying out the renormalization of F.
\end{enumerate}

The third point suggests a clean  explanation of  an  old puzzle of pQCD: the absence
of   even zetas in  the Adler function of pQCD at order $\alpha_s^3$, $D_{(3)}(q^2)$.
Indeed, the function at this order is 

(i) a {\em finite} combination of four-loop p-integrals; 

(ii) the corresponding renormalization is done with   the help of charge coupling renormalization which does not
      depend on any zetas at the order required.

As a direct consequence of (ii)  the function $D_{(3)}(q^2)$ {\bf
should not} depend on $\zeta_4$, $\zeta_6$ and $\zeta_3 \zeta_4$.
This is indeed the case \cite{GorKatLar91:R(s):4l,SurSam91}!  At
first glance, this explanation of this old puzzle generates another
one: why $D_{(3)}(q^2)$ does not include terms proportional odd zetas with
weight larger than five (which is expected in general for any
combination of four-loop p-integrals)? No, it is not a puzzle since
long.  It is a  well-known fact\footnote{First, probably understood on
the example of the three-loop ${\cal O}(\alpha_s^2)$ Adler function
\cite{Chetyrkin:1979bj}.} that the (L+1) loop Adler function (in massless QCD) could be
completely expressed through L-loop p-integrals.

Exactly the same reason explains the absence of even zetas
in  the four-loop contribution   to
the Bjorken sum rule \cite{PhysRevLett.104.132004}.

 The problem of why the five-loop   ${\cal O}(\alpha_s^4)$ Adler function is also free
from even zetas should be possible to  solve by extending the above reasoning  by one loop higher. 
The "only"  missing  ingredients  --- a property of five-loop master p-integrals analogous to 
\re{hat_zetas}. We hope to come  back to  the subject in future.

\subsection{Three loops  \label{evenz:3}}

In fact,  Theorem 4  was  proven for the three-loop p-integrals
in an early work  by Broadhurst \cite{Broadhurst:1999xk}  with the help of  essentially equivalent
(though, to our opinion, somewhat more complicated)  considerations. 
This is certainly enough to explain the absence of $\zeta_4$ in 
the $\alpha_s^2$ contribution to   Adler function \cite{Chetyrkin:1979bj}  and in the 
$\alpha_s^3$ ones to the deep inelastic sum rules found in \cite{Larin:1990zw,Larin:1991tj}. 
This is  because all these  quantities are naturally expressed through some  combinations 
of   {\em three-loop} p-integrals with purely rational coefficients.

Note that, the three-loop version of Theorem 4 \cite {Broadhurst:1999xk} is
not enough to explain the absence of even zetas from the $\alpha_s^3$
contribution to Adler function and from the four-loop result for the
QCD $\beta$-function \cite{vanRitbergen:1997va,Czakon:2004bu}. The
problem is that, to the best of our knowledge, it is not known whether
one could find a representation, say, the four-loop contribution to
the QCD $\beta$-function in term of a finite combination of the
three-loop p-integrals with coefficients {\em free} from any zetas. The
same is true for the Adler function.

On the other hand, we do agree with \cite{Broadhurst:1999xk} that the
four-loop QED $\beta$-function should be free from $\zeta_4$ (in
agreement with explicit calculations of \cite{Gorishnii:1991hw}) as it can be 
expressed via  a {\em finite} combination (with purely rational  coefficients) of three-loop p-integrals
\cite{Johnson:1973pd}.

\section{Discussion \label{discussion}}

There are various points deserving further  discussion in connection with  the algorithm of evaluation of MI's
elaborated in sections \ref{3loop} and \ref{4loop}.

\begin{itemize}

\item 
The results discussed in the present paper have been indispensable for
the long-term project of computing the cross section of $e^+ e^-$
annihilation into hadrons at order $\alpha_s^4$ in QCD
\cite{Baikov:2008jh,PhysRevLett.104.132004}.  While they were  first obtained
in 2003, their publication had been postponed in favour of the
faster completion of the main project.

\item 
Recently, a definite class of massless p-integrals was proven to be
expressible in terms of the multiple zeta values for all orders of
expansions in $D-4$, and a direct method of their evaluation was
suggested (\cite{Brown:2008um,Brown:2009ta}).

In our case (four-loop p-integrals) it predicts that the result is expressible
through $\zeta_n$ up to $n=7$, as  confirmed by our calculations.
Unfortunately this method in its present form is applicable only for
integrals with number of internal lines equal to doubled number of
loops plus one, so the most complicated  four-loop MI's seem to be unreachable.

\item The heavy use of identities between Feynman integrals coming,
eventually, from IBP relations is not a unique feature of our approach
to the evaluation of MI's.  It is of interest, that three other, quite
different  and in a sense  more general approaches  would also be impossible 
without  intensive  use of the reduction of Feynman integrals to masters. We mean (i) the method
of differential equations\footnote{ Starting from early works 
\cite{Kotikov:1990kg,Kotikov:1991hm,Kotikov:1991pm,Kotikov:1990zs,Kotikov:1991mg,Kotikov:1990zk,Remiddi:1997ny,%
Caffo:1998du,Caffo:1998yd} the method has developed into quite  a powerful technique. For its  modern status
and further references, see the  review \cite{Argeri:2007up}.}, 
(ii) the use of difference
equations \cite{Laporta:2001dd,Laporta:2000dc} and, at last, very new method  \cite{Kirilin:2008hu,Lee:2009dh,Lee:2010cg} 
based on recurrence equations with respect to the space-time dimension D \cite{Tarasov:2000sf}.

Finally, we want to mention two  popular and powerful methods of evaluation of MI's 
which do not use  directly the IBP reduction. The first approach is  based on the Mellin-Barn  representation.
The  early applications of Mellin integrals to evaluation of FI's  
were performed  in pioneering works \cite{Usyukina:1989yg,Boos:1990rg}. 
Currently it  is an actively developing field, for  a review see, e.g.  \cite{Smirnov:2004ym,Smirnov:2006ry}.

The second method --- the so-called sector decomposition --- was originally used as a convenient theoretical tool 
for the analysis of convergence  of FI's \cite{Hepp:1966eg,Speer:1975dc}. First applications of sector decomposition   
for evaluation of FI's  were considered in \cite{Binoth:2000ps,Binoth:2003ak,Binoth:2004jv}. 
The current status of the method can  be found in a review \cite{Heinrich:2008si}.

\item 
  While
the tests of our results described in subsections \ref{test:primitive} and \ref{test:simple}   leave no room for doubt as for the
cases of {\em trivial} and {\em simple} groups of MI's, it is {\em not} true for
the most difficult group of {\em complicated} integrals: for this family  of 
 thirteen MI's  only three  had been directly checked in an  independent way (see subsection  \ref{test:complicated}).  
It means that if a master integral from the remaining ten integrals   were assigned
a wrong value, it would  {\em change}  in  all probability {\em all} physical results obtained 
with the  use of these  MI's  since 2004.

Let us, therefore,  discuss a little bit further the 
important issue of the correctness of these  MI's.
The method of computing  master p-integrals described in sections \ref{3loop} and \ref{4loop} heavily 
uses both the GaC symmetry {\em and} the procedure of reduction of four-loop  p-integrals.
The latter is the  most complicated part of all  the  calculation 
as it  requires, first,  careful computer algebra  programming  and, second, 
large-scale calculations.  
Thus,  an independent check of the ten remaining  most 
complicated  MI's would also provide us with a quite strong, though non-direct, evidence for 
the correctness of the  reduction algorithm  we use and its FORM implementation.

Fortunately, such an independent check of {\em all} complicated integrals
has been  very recently performed in 
\cite{Tentyukov:masters:10} with the use of the  sector decomposition, where not only all results
have been (numerically) confirmed with better than 1\promille \ 
accuracy\footnote{ We mean the accuracy for the most complicated last
${\cal O}(\ep^{p_i})$ term in comparison  with  the exact results listed in eqs.~(\ref{m51}-\ref{m43}), 
the accuracy of simpler terms of order
$\ep^i$ with ${-4 \leq i< p_i})$ is significantly higher. The typical
accuracy of the ${\cal O}(\ep^{p_i+1})$ term (which is necessary only
in evaluation of {\em five}-loop master p-integrals)  was also about 1\promille. \  One should keep in mind that
the latter accuracy is an estimate given  by the MC-integrator and as such it is not always reliable.  } 
but also one extra term in these $\ep$-expansions have been
computed.

\end{itemize}

\section{Summary and Conclusions    \label{conclusions}   }

In this work we have presented an algorithm for the  analytical evaluation
of all master integrals which appear in the process of reduction of
massless dimensionally regulated Feynman integrals with one external
momentum (p-integrals).  The algorithm is based on  the glue-and-cut
symmetry \cite{Chetyrkin:1981qh} which is an unique and very specific
property of such integrals   valid irrespectively of their
complexity (number of loops).  In addition to  the symmetry the  algorithm
heavily uses the  reduction procedure. 

It has been demonstrated that the algorithm works flawlessly for the
case of the three-loop p-integrals (successfully reproducing well-known
thirty years old results of \cite{Chetyrkin:1980pr}) and four-loop
p-integrals. In the latter case it produces explicit analytical results
for all  master integrals, major part of which are  new.

Together with   Theorem 2 and the $1/D$ method of reduction of p-integrals \cite{Baikov:2005nv,Baikov:2007zza}
the  algorithm  guaranties that the UV counterterm  of any five-loop  diagram can be calculated 
within the MS-scheme in terms of rational numbers, $\zeta_3$, $\zeta_4$, $\zeta_5$, $\zeta_6$ and
$ \zeta_7$. This implies the analytical calculability of the $\beta$-functions and anomalous dimensions
of fields and composite operators in an arbitrary model at  the five-loop level.

\noindent
{\em Acknowledgments.}

This work was supported by the Deutsche Forschungsgemeinschaft in the
Sonderforschungsbereich/Transregio SFB/TR-9 ``Computational Particle
Physics'' and  by RFBR grants  08-02-01451, 10-02-00525.

We are very grateful to J.H.~K\"uhn for  his long-term encouragement, support and patience.

We thank  D.~Broadhurst, M.~Czakon, A.~Grozin, S.~Moch,  V.A.~Smirnov and  M.~Tentyukov   
for fruitful discussions, attentive reading of the manuscript and good advice.
K.G.~Chetyrkin is grateful to A.~Czarnecki for  a discussion of hystorical aspects of the work.  
                
K.G.~Chetyrkin should also thank  A.L.~Kataev and  F.V.~Tkachov for a fruitful collaboration 
during their Diploma studies 
which started the exploration  of the wonderful land of  massless propagator
like FI's \cite{Chetyrkin:1979bj}.  The present work is  dedicated to the memory of his  late
friend and  colleague  \mbox{S.G.~Gorishny}, whose contribution was decisive in another important
application of the idea of glueing ---  the derivation of the Wilson short-distance expansion
in the minimally subtracted  scheme \cite{Chetyrkin:1982zq} 
and, as a logical consequence, the creation of the
method of projectors~\cite{Gorishnii:1983su}.

\providecommand{\href}[2]{#2}\begingroup\raggedright\endgroup

\end{document}